\documentclass[aps,twocolumn,nofootinbib,groupedaddress,amsfonts,floatfix]{revtex4-1} 
\pdfoutput=1
\usepackage{graphicx,amsmath,amssymb,amstext}
\usepackage{amssymb,amsbsy,amsfonts,amsthm,color}

\usepackage{epsfig}
\usepackage{graphicx}
\usepackage{subfigure}
\usepackage{hyperref}

\graphicspath{{Figures/}}

\usepackage{color}
\newcommand{\Beq}{\begin{eqnarray}}
\newcommand{\Eeq}{\end{eqnarray}}

\newcommand{\eqn}[1]{Eqn. (\ref{#1})}

\def\lsim{\mathrel {\vcenter {\baselineskip 0pt \kern 0pt \hbox{$<$} \kern 0pt \hbox{$\sim$} }}}

\newcommand{\mpl}{M_{\mbox{\tiny Pl}}}

\def\gsim{\mathrel {\vcenter {\baselineskip 0pt \kern 0pt \hbox{$>$} \kern 0pt \hbox{$\sim$} }}}

\newcommand{\grchombo}{\mathtt{GRChombo}}

\begin{document}
{\hfill KCL-PH-TH/2017-65}
\vskip .5cm

\title{Robustness of Inflation to Large Tensor Perturbations}

\author{Katy Clough${}$}
\email{katy.clough@phys.uni-goettingen.de}
\affiliation{${}$Instit{\"u}t f{\"u}r Astrophysik, Georg-August Universit{\"a}t, Friedrich-Hund-Platz 1, D-37077 G{\"o}ttingen, Germany}
\author{Raphael Flauger${}$}
\email{flauger@physics.ucsd.edu}
\affiliation{${}$ Center for Astrophysics and Space Sciences, University of California, San Diego, 9500 Gilman Drive 0319 CA La Jolla 92093, USA}
\author{Eugene A. Lim${}$}
\email{eugene.a.lim@gmail.com}
\affiliation{${}$Theoretical Particle Physics and Cosmology Group, Physics Department, Kings College London, Strand, London WC2R 2LS, United Kingdom}

\begin{abstract}
Extending our previous work on the robustness of inflation to perturbations in the scalar field, we investigate the effects of perturbations in the transverse traceless part of the extrinsic curvature on the evolution of an inhomogeneous inflaton field. Focusing on small field models, we show that these additional metric inhomogeneities initially reduce the total number of $e$-folds as the amplitude increases, but that the reduction saturates and even reverses above a certain amplitude. We present an argument that this is due to the presence of a large initial Hubble friction when metric perturbations are large.
\end{abstract}

\pacs{}
\maketitle
\section{Introduction}
\label{sec-intro}

Inflation \cite{Guth:1980zm,Linde:1981mu,Albrecht:1982wi,Starobinsky:1980te} was proposed as a solution to problems in standard Big Bang theory such as the horizon and flatness problem. The solution relies on the ability of inflation to ``inflate away'' initial inhomogeneities, dynamically generating a homogeneous and isotropic Universe with a nearly scale-invariant power spectrum of primordial perturbations consistent with observations. However, this is unlikely to succeed for all possible initial conditions, and one may ask what the requirements on the initial data are for inflation to succeed. 

If general relativity coupled to a scalar field provides a suitable description of our universe all the way to Planckian energy densities the requirements are minimal~\cite{Linde:1984ir,Linde:1985ub,Linde:2014nna}. However, both observations of the cosmic microwave background~\cite{Ade:2015lrj} and theoretical considerations motivate a study of initial data that leads to successful inflation assuming that this description only becomes appropriate at sub-Planckian energy scales.  

The question of ``initial conditions'' for inflation in various setting has been studied extensively using perturbative and dynamical systems approaches, and there are many analytic and semi-analytic~\cite{Gibbons:1977mu,Hawking:1981fz,Wald:1983ky,Starobinsky:1982mr,Barrow:1984zz,Albrecht:1984qt,Barrow:1985,Gibbons:1986xk,Jensen:1986nf,Hawking:1987bi,Penrose:1988mg,Muller:1989rp,Kitada:1991ih,Kitada:1992uh,Bruni:1994cv,Maleknejad:2012as,Gibbons:2006pa,Boucher:2011zj,Bruni:2001pc,Muller:1987hp,Barrow:1989wp,Bicak:1997ne,Capozziello:1998dq,Vachaspati:1998dy,Barrow:1987ia,Barrow:1986yf,Polyakov:2009nq,Marolf:2010nz,Tsamis:1992sx,Brandenberger:2002sk,Geshnizjani:2003cn,Marozzi:2012tp,Brandenberger:1990wu,Carroll:2010aj,Corichi:2010zp,Schiffrin:2012zf,Remmen:2013eja,Corichi:2013kua,Mukhanov:2014uwa,Remmen:2014mia,Berezhiani:2015ola,Kleban:2016sqm} as well as numerical studies~\cite{Albrecht:1985yf,Albrecht:1986pi,KurkiSuonio:1987pq,Feldman:1989hh,Brandenberger:1988ir,Goldwirth:1989pr,Goldwirth:1989vz,Brandenberger:1990xu,Laguna:1991zs,Goldwirth:1991rj,KurkiSuonio:1993fg,Easther:2014zga,East:2015ggf,Braden:2016tjn,Alho:2011zz,  Alho:2013vva} (see~\cite{Brandenberger:2016uzh} for a short review). Recently it has become possible to use numerical relativity codes to evolve different initial configurations in the time domain ~\cite{East:2015ggf, Clough:2016ymm}.
The use of numerical relativity, in which the full Einstein equations are evolved, permits the exploration of configurations beyond the perturbative regime, and also gives useful insights into the mechanisms by which the slow roll period can be disrupted.

In our previous paper~\cite{Clough:2016ymm}, we studied the robustness of large and small field inflationary models to inhomogeneities in the inflaton field. We allowed inhomogeneity in the conformal factor and the trace of the extrinsic curvature, but did not allow inhomogeneities in the traceless part of the extrinsic curvature.

The transverse traceless part of the extrinsic curvature heuristically describes a background of gravitational waves. More precisely, in the limit in which the perturbations in the transverse traceless part are small, they correspond to the two gravitational wave polarization modes. When their amplitudes are large, their energy density does not scale like gravitational radiation and we use the term gravitational wave background loosely to include large inhomogeneities. 

Setting these perturbations to zero on the initial slice as in~\cite{East:2015ggf, Clough:2016ymm} is natural in most applications of numerical relativity, in particular for black hole mergers, because one is interested in the gravitational waves produced by the merger. However, for inflation one may expect a gravitational wave background to be present. Setting this background to zero is a strong assumption only made in~\cite{East:2015ggf,Clough:2016ymm} to simplify the computations. In this paper we study more general inhomogeneities with non-zero transverse traceless part of the extrinsic curvature. 

\begin{figure*}
\begin{center}
\includegraphics[width=13.2cm]{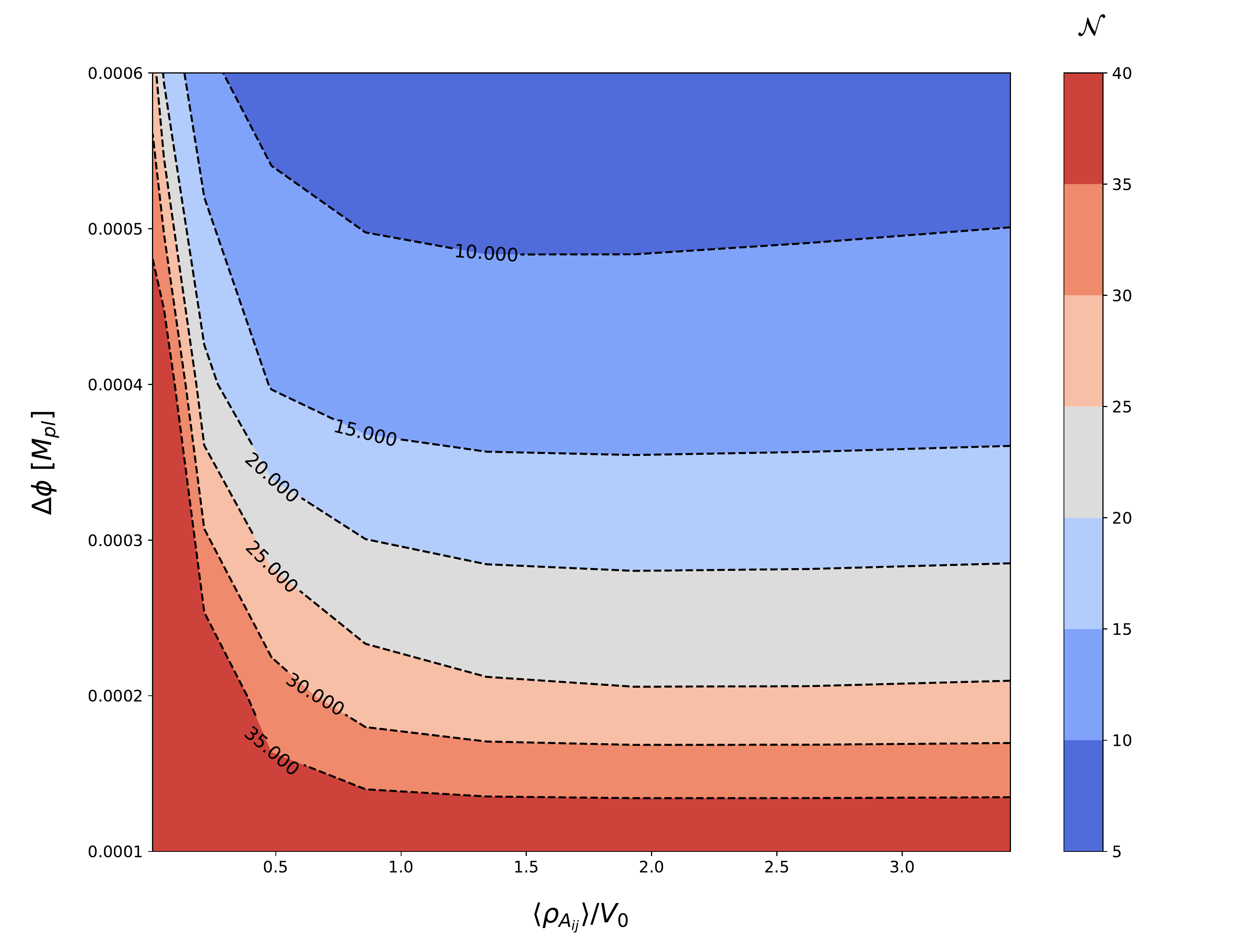}
\caption{Number of $e$-folds of inflation versus $\langle \rho_{A_{ij}} \rangle/V_0$ (relating to the amplitude of the tensor perturbations') and $\Delta\phi$ (the amplitude of fluctuations in the inflaton field). The decrease in $e$-folds with increasing $\Delta\phi$ is consistent with our previous work, showing a power law decrease. The amount of inflation initially decreases as the gravitational wave fluctuations are increased, but the decrease levels off and even recovers slightly at $\langle \rho_{A_{ij}} \rangle/V_0 \sim 1$. Above $\langle \rho_{A_{ij}} \rangle/V_0 \sim 4$ the fluctuations reach a level at which they undergo gravitational collapse, and we are not able to evolve the simulations to the end of inflation.
\label{fig:Efolds2D}}
\end{center}
\end{figure*}

We will largely focus on small field models, which are subject to the failure mode in which the field ``falls off the potential hill'', and into minimum, when the fluctuations become too large. We consider horizon scale modes in the  scalar field perturbations as these were found to be the most problematic for the onset of inflation in our previous work~\cite{Clough:2016ymm}.

Our key results are shown in figure \ref{fig:Efolds2D} and are summarised here for convenience
\begin{itemize}
\item{{\it Inhomogeneities in the transverse traceless part of the extrinsic curvature lead to a reduction in the number of $e$-folds compared to the case in which the inhomogeneities were confined to the inflaton field and the conformal factor.}  This behaviour is illustrated in figure \ref{fig:Efolds2D}.}
\item{{\it Scalar field inhomogeneities are required for failure.} Adding a tensor background to a spatially homogeneous inflaton field did not lead to failure even for high amplitudes of tensor perturbations.}
\item{{\it There exists an upper bound on the effects of tensor fluctuations on the robustness of inflation.} For energy densities in these inhomogeneities above the inflationary energy density $V_0$, the reduction in the number of $e$-folds saturates and even reverses as the amplitude of the fluctuations increases. We propose that this is mainly due to the increased Hubble friction, as a result of which the oscillations of the inflaton field are ``damped'' by the effective energy density of the metric perturbations. We propose that large tensor fluctuations could therefore mitigate the disruptive effect of a small non zero initial value of $\dot\phi$.}
\item{{\it At higher amplitudes, increases in the extrinsic curvature fluctuations result in local collapse.} For energy densities in the homogeneities above $\sim 4 V_0$ regions of the spacetime begin to collapse. We expect black holes to form from the collapses we observed but we were not able to continue the simulations long enough to observe the apparent horizons forming. We do not expect these collapses to impede slow roll inflation in the remaining spacetime.}
\item{{\it The failure of the critical point can be simply modelled.} We use insights from our simulations to propose a model for the failure of the critical point, which allows us to recreate our results and explore the behaviour of larger metric fluctuations. The model could be applied to any additional energy density component with a known scaling versus $e$-folds.}
\item{{\it Large field inflation is robust to large tensor perturbations. } We confirmed that large field models continue to inflate in the presence of tensor inhomogeneities.}
\end{itemize}

This paper is organised as follows. In Section \ref{sect:method} we present the theory and methodology of our approach with reference to our previous work, with some of the code detail contained in Appendix \ref{sect:AppendixA} to keep the discussion clear. In Section \ref{sect:results}, we present the results, and discuss their interpretation.

\section{Theory and Methodology} \label{sect:method}

As in \cite{Clough:2016ymm}, we consider a single inflaton field with a canonical kinetic term
\begin{equation}
L_{\phi} = -\frac{1}{2}g^{\mu\nu}\partial_\mu \phi\partial_\nu\phi - V(\phi) \label{eqn:singlefield}\,.
\end{equation}
We choose a slow roll potential, illustrated in Figure \ref{fig:SF}, with initial conditions so that inflation would last 100 $e$-folds or more the homogeneous case. Inhomogeneities will in general reduce this number, and we will investigate the reduction in the number of $e$-folds. 

We decompose the metric into a spatial metric $\gamma_{ij}$, lapse, and shift vector using the ADM formalism~\cite{Arnowitt:1962hi} (see Appendix \ref{sect:AppendixA} for further details). 
 
The initial conditions are subject to the usual Hamiltonian and momentum constraint equations \eqn{eqn:Ham} and \eqn{eqn:Mom}.
The constraint equations are coupled elliptic partial differential equations and are in general challenging to solve. To make progress, we study restricted classes of initial conditions for which these equations simplify and make them progressively more general. 
In our previous work \cite{Clough:2016ymm}, we restricted ourselves to scalar field profiles with spatially constant velocity and vanishing transverse traceless part of the extrinsic curvature on the initial slice.

In this paper, we also make the simplifying assumption that the scalar field profiles have constant velocity on the initial slice, but we free the transverse traceless part of the extrinsic curvature. These inhomogeneities are not sourced by the matter content, but rather represent dynamical fluctuations in the metric.

In particular, we emphasise that there is no \emph{a priori} reason why these degrees of freedom should be vanishing. These metric choices are described in section \ref{sect:ICMetric} below. The matter configuration is described in section \ref{sect:ICMatter}.

We evolve these initial conditions forward in time using the code GRChombo \cite{Clough:2015sqa}, using the BSSN formalism of Numerical Relativity (NR). Key details on the evolution setup are provided in section \ref{sect:Evolution}, and some, particularly generic numerical details of the code, are provided in the Appendix \ref{sect:AppendixA}. \footnote{Note that we now use a conformal factor defined by $\chi ~ \gamma_{ij} = \tilde{\gamma}_{ij}$. We have updated the arXiv version of \cite{Clough:2016ymm} to agree to the new convention, since it is more consistent with the naming in other literature in NR. We apologise for any confusion this may cause.}

\subsection{Initial Conditions - matter}
\label{sect:ICMatter}

We impose simple inhomogeneous initial conditions on the matter sector of the form\footnote{We limit ourselves to a single wavelength because we saw in our previous work that inflation is least robust to the longest wavelength modes.} 
\begin{multline}
\phi(t = 0,{\bf x}) = \phi_0 \\ + \Delta \phi \left( \cos{\frac{2 \pi x}{L}} + \cos{\frac{2 \pi y}{L}} + \cos{\frac{2\pi z}{L}} \right) \label{eqn:phi}\,,
\end{multline}
and unless otherwise stated
\begin{eqnarray} \label{eqn:phidot}
\frac{\partial \phi(t = 0,{\bf x})}{\partial t} = 0 \,,
\end{eqnarray}
where ${\bf x}$ is the spatial coordinate of a foliation labeled by the time coordinate $t$, and $\Delta \phi$ is a measure of the amplitude of the initial inhomogeneities. The maximal total amplitude of the fluctuations about $\phi_0$ in this case is $3\Delta \phi$. 
The value $\phi_0$ is chosen such that we have 100 $e$-folds of inflation in the absence of inhomogeneities. 

\begin{figure}
\begin{center}
\includegraphics[width=8cm]{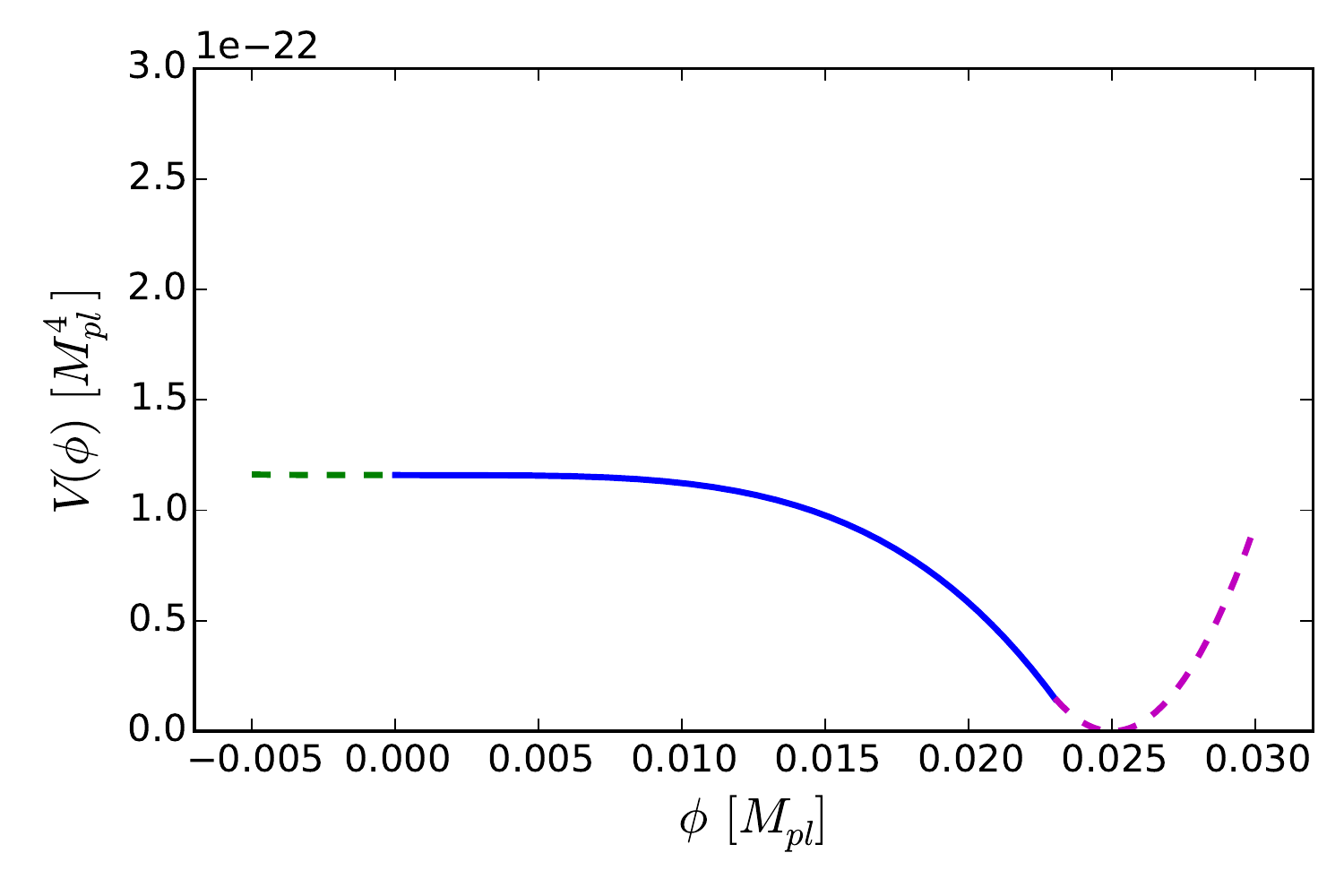}
\caption{Our toy slow roll potential, which corresponds to ``small field'', low-scale inflation.
\label{fig:SF}}
\end{center}
\end{figure}

We set $L$ to be the length of the simulation domain, and use periodic boundary conditions to simulate a space composed of periodic fluctuations of this length and amplitude. $L$ is chosen to be the Hubble length in the absence of inhomogeneities ($\Delta \phi=0$), that is
\begin{eqnarray}
L \equiv H_0^{-1} = \frac{3 \mpl}{\sqrt{24 \pi V_0}} \label{eqn:Lformula}\,,
\end{eqnarray}
where
\begin{equation}
V_0 \equiv V(\phi_0)
\end{equation}
is the potential energy density when $\phi(t=0,\mathbf{x})=\phi_0$.
In the small field case, which we mainly investigate, we choose a potential with an extended flat plateau in one direction, as shown in figure \ref{fig:SF}. As in~\cite{Clough:2016ymm}, we model the inflationary potential as
\begin{equation}
V(\phi)=
\begin{cases}
V_0& \phi<0\\
V_0 \left(1 - \left(\frac{\phi}{\mu}\right)^4\right) & 0<\phi<0.023 \mpl\\
m^2 (\phi - \phi_*)^2 & \phi > 0.023 \mpl
\end{cases}
\end{equation}
with $\mu = 0.0238\mpl$, $V_0 = 1.15949 \times 10^{-22} \mpl^{4}$, \mbox{$m^2 = 3.75 \times 10^{-18} \mpl^2$} and $\phi_* = 0.025\mpl$. The Hubble rate during inflation for this choice of parameters is $H_{\rm inf} = 3.125 \times 10^{-11} \mpl$, and for a (homogeneous) initial value of the field of $\phi_0 = 0.001\mpl$, these values would result in 100 $e$-folds of inflation.

In the large field case, we also use the same model and choice of parameters as in \cite{Clough:2016ymm}, that is
\begin{eqnarray}
V(\phi) = m^2 \phi^2\,, 
\end{eqnarray}
with $m=1.07967 \times 10^{-7} \mpl$. For an initial value of the field of $\phi_0 = 4\mpl$, this results in 100 $e$-folds of inflation, a scale of inflation $H_{\rm inf} = 1.25 \times 10^{-6} \mpl$, a scalar perturbation amplitude $\Delta_R = 10^{-5}$ and scalar spectral index $n_s \approx 0.97$ for modes that exit the horizon 60 $e$-folds before the end of inflation.

In summary, our model of initial matter inhomogeneities depends on the amplitude of inhomogeneities $\Delta \phi$, and the potential $V(\phi_0)$. The potential $V(\phi_0)$ sets the inflationary Hubble scale, which in turn sets the wavelength of the perturbations. We also consider a case where $\dot\phi$ is a small non zero constant on the initial spatial slice, which is described further below. 

\subsection{Initial Conditions - metric}
\label{sect:ICMetric}

We define the conformal metric $\tilde{\gamma}_{ij} = \chi^{-1}\gamma_{ij}$ where $\chi$ is a scalar conformal factor.  In \cite{Clough:2016ymm}, we made the simplifying assumption that the traceless part of the extrinsic curvature was zero everywhere on the initial slice
\begin{equation}
A_{ij} \equiv K_{ij} -\frac{1}{3}\gamma_{ij} K=0\,,\label{eqn:tracelessK}
\end{equation}
where $K_{ij}={\cal L}_{{\bf n}}\gamma_{ij}$ is the extrinsic curvature and ${\bf n}$ the normal vector to the spatial slices. In addition, we assumed that the metric on the initial slice was conformally flat
\begin{equation}
\tilde{\gamma}_{ij} = \delta_{ij}\,.  \label{eqn:conformalcond}
\end{equation}

In this work we consider $A_{ij} \neq 0$ but restrict ourselves to variation of $K$ that allows us to set the longitudinal part of
\begin{equation}
\bar A^{ij} = \chi^{-5/2} {A}^{ij}\,,
\end{equation}
to zero.
Note that this conformal version is denoted by a bar to differentiate it from the version used in the BSSN conformal decomposition,
\begin{equation}
A_{ij} = \chi \tilde{A}_{ij} ~ ,
\end{equation}
which is denoted by a tilde over the $A$.

To see that this is possible notice that with this alternative conformal decomposition, the momentum constraint can be written as
\begin{equation}
\tilde{D}_j \bar{A}^{ij}  -\frac{2}{3}\chi^{-3/2}\tilde{D}^j K = 8\pi G \chi^{-5/2} S^i. \label{eqn:MomCon1}
\end{equation}
Since the conformal metric $\tilde{\gamma}_{ij} = \delta_{ij}$, the covariant derivative $\tilde D$ reduces to a normal partial derivative. 

For spatially constant $K$, and $\dot\phi=0$, the momentum density vanishes, $S^i=0$, and we see that the longitudinal part of $\bar{A}^{ij}$ vanishes. More generally, for a constant scalar field velocity on the initial slice, the longitudinal part of $\bar{A}^{ij}$ vanishes as long as the spatial variation of $K$ compensates for the spatial variation of the scalar field as in~\cite{Clough:2016ymm}. 

Under these assumptions, the momentum constraint reduces to 
\begin{equation}
\partial_j \bar{A}^{ij} =0. \label{eqn:MomCon2}
\end{equation}
Setting the longitudinal part to zero obviously provides a solution, but let us also show that on a torus, as we consider here, this is the only solution for our choice of initial conditions.  

To see this, we decompose $\bar A_{ij}$ into a longitudinal and transverse part
\begin{equation}
\bar{A}^{ij} = \bar A^{ij}_{TT} + \bar A^{ij}_L. \label{eqn:decomposebarAij}
\end{equation}
where the transverse part $A_{TT}^{ij}$ satisfies
\begin{equation}
 \partial_j \bar{A}^{ij}_{TT} = 0. \label{eqn:AijTT}
\end{equation}
We can then express the longitudinal part in terms of a vector $W^i$ 
\begin{equation}
\bar{A}^{ij}_L = \partial^i W^j +  \partial^j W^i - \frac{2}{3} \delta^{ij}  \partial_k W^k ~. \label{eqn:AijL}
\end{equation}

The momentum constraint then implies 
\begin{equation}
\partial^j \partial_j W^i + \frac{1}{3} \partial^i \partial_j W^j = 0. \label{eqn:MomCon4}
\end{equation}
On a torus, the only solution to this equation are constant vector fields so that $\bar{A}^{ij}_L=0$.\footnote{In more general spacetimes non-trivial solutions exist. In asymptotically flat spacetimes the solutions are harmonic, as was used in \cite{Shibata:1995we}. It would be interesting to consider the effect of such perturbations in more general spacetimes.} 

Since we are working on a torus, we can Fourier decompose $\bar A^{ij}_{TT}$. This reduces finding a transverse traceless tensor to a problem in linear algebra. More specifically, for the Fourier expansion
\begin{equation}
\bar A^{ij}_{TT}=\sum_{n_1,n_2,n_3}c_{(n_1,n_2,n_3)}^{ij}\exp\left[\sum_k \frac{2\pi i n_k x^k}{L}\right]\,,
\end{equation}
the conditions that $\bar{A}^{ij}_{TT}$ be transverse and traceless imply
\begin{equation}
\sum_i n_i c_{(n_1,n_2,n_3)}^{ij}=0\quad\text{and}\quad\sum_i c_{(n_1,n_2,n_3)}^{ii}=0\,.\\
\end{equation}
We see that these equations impose four conditions on six independent matrix elements, leaving us with two degrees of freedom. In the linear regime, these are, of course, the two polarization states of the graviton. To ensure $\bar A^{ij}_{TT}$ is real we must choose
\begin{equation}
c_{(-n_1,-n_2,-n_3)}^{ij}={c_{(n_1,n_2,n_3)}^{*\,ij}}\,.
\end{equation}

These equations are straightforward to solve for any choice of $n_1$, $n_2$ and $n_3$. Given a choice of $\bar A^{ij}$, all quantities appearing in the Hamiltonian constraint
\begin{multline}
\tilde{D}^2\chi -\frac{5}{4 \chi}\tilde{\gamma}^{ij}\tilde{D}_i\chi \tilde{D}_j\chi \\ + \frac{\chi \tilde{R}}{2} + \frac{K^2}{3} - \frac{1}{2} \tilde{A}_{ij}\tilde{A}^{ij} = 8\pi G \rho \label{eqn:HamCon}\,,
\end{multline}
with the exception of the conformal factor $\chi$ are known, and we can solve for $\chi$ numerically. (We do this by relaxation from a trial solution.)\footnote{Note that one must reconstruct the value of the evolution variable $\tilde A_{ij}$ which appears in this expression from the chosen $\bar A_{ij}$ and the conformal factor (a procedure which must form part of the relaxation, as the conformal factor is changing). In addition, as was noted previously, the relaxation of $\chi$ could result in a profile which is inconsistent with the initial value of $K$ per \eqn{eqn:Kavg} which was calculated assuming $\chi=1$. In principle we may need to adjust the value of $K$ to better satisfy the constraint, but in practise this is rarely necessary.}

For compact spatial slices as we assume here, the value of $K$ is not independent, but is determined by the Hamiltonian constraint~\eqn{eqn:HamCon}
\begin{equation}
K \simeq - \sqrt{\langle 24 \pi G \rho + \frac{3}{2} \tilde A_{ij} \tilde A^{ij}}  \rangle  \,, \label{eqn:Kavg}
\end{equation}
with
\begin{equation}
\rho = \frac{1}{2}(\partial_i \phi)^2 + V(\phi)\,, \label{eqn:rhototal}
\end{equation}
where $\langle X \rangle$ denotes the average of $X$ over the spatial spatial slice. 
Here the `$\simeq$' sign indicates that the exact expression involves the values of $\chi$ which are not yet known. For our choice of initial conditions, it can be approximated by unity. We then solve (by relaxation) the Hamiltonian constraint for the spatially varying values of $\chi$, and check that this yields a consistent solution. To improve the solution, this can be done iteratively, but in the cases we consider this was not necessary as it did not significantly change the solution obtained - for further details see figure \ref{fig:Convergence_Ham} in Appendix \ref{iteration}.

As for the scalar perturbations, one expects short wavelength modes of the transverse traceless part of the extrinsic curvature to behave like radiation, and one expects modes with wave numbers comparable to the local expansion rate to be the most problematic for the onset of inflation. So in practice, we consider a simple superposition of modes with wave numbers $(n,0,0)$, $(0,n,0)$ and $(0,0,n)$ with equal amplitudes of the form
\begin{equation}\label{eq:Aijf}
\bar A^{ij}_{TT} = 
  \left[ {\begin{array}{ccc}
   0    & f(z) & f(y)\\
   f(z) & 0    & f(x)\\
   f(y) & f(x) & 0   \\
  \end{array} } \right]
\end{equation}
with
\begin{equation}
f(t = 0, x) =\Delta A  \cos\left(\frac{2 \pi nx}{L} \right) \label{eqn:A_ij_modes}\,.
\end{equation}
Here $\Delta A$ sets the amplitude of the fluctuations, $n$ sets the wavenumber of the fluctuations relative to the inflationary Hubble scale $H_0$ (see \eqn{eqn:Lformula}, and we consider the cases $n=1$ and $n=6$. In each case, we consider a single mode, with a constant spatial $K$ and $\Delta A$. Note that since the actual Hubble radius will be smaller due to the presence of energy density from inhomogeneities, $n=1$ (and possibly $n=2$)  modes are superhorizon.

In addition, we consider the case of a superposition of modes with wave vectors $(1,3,5)$, $(2,4,4)$, and $(4,3,3)$. Each mode is normalized to carry the same energy density as an individual mode in equations~(\ref{eq:Aijf}),~(\ref{eqn:A_ij_modes}). In particular, for this normalization the average energy density $\langle \rho_{A_{ij}} \rangle$ carried by the superposition of the three modes for a given $\Delta A$ is the same as for the initial conditions in equations~(\ref{eq:Aijf}),~(\ref{eqn:A_ij_modes}). We will refer to this superposition as anisotropic and the superposition of wave vectors $(n,0,0)$, $(0,n,0)$, and $(0,0,n)$ as isotropic initial conditions, respectively.

Finally, we consider whether the gravitational wave background can mitigate the instability of small field models to an initial non zero value of $\dot\phi\neq 0$. While in general this requires us to solve the momentum constraints for the initial conditions, it turns out that there exists a special case where these constraints are trivially solved when $\dot \phi=C/(12\pi) \mpl^2$ and the spatial variation in $K = C \phi + K_0$. Here, we choose $C=0.0005$ and the value of $K_0$ is chosen as a constant value which permits us to satisfy the Hamiltonian constraint with periodic boundary conditions (see \cite{Clough:2016ymm} for further details on this approach). We report on the results of this case in section \ref{sec-interpretation}.

\subsection{Evolution of the initial conditions}
\label{sect:Evolution}

We use the numerical relativity package $\grchombo$~\cite{Clough:2015sqa} for the time evolution. Full details of the numerical implementation, evolution equations and gauge choices are provided in Appendix \ref{sect:AppendixA}, along with plots showing the convergence and bounded constraint violation in typical simulations.

We ran three cases for the scalar perturbations, with $\Delta \phi=0.0006 \mpl$, $\Delta \phi=0.0004 \mpl$ and $\Delta \phi=0.0002 \mpl$, scanning over the values for $\Delta A$. These cases would have resulted in over 40 $e$-folds of inflation (the maximum we can follow in one simulation), in the absence of any perturbations in $A_{ij}$.  For each case we gradually increased the value of $\Delta A$ from $0$ to $2 \times 10^{-10} \mpl$ and recorded the number of $e$-folds at the point of failure. Above this level of $\Delta A$ the spacetime undergoes gravitational collapse due to the presence of the perturbations in the metric, and we were not able to evolve the simulations until inflation ended.

It is convenient to define the following local energy density to encapsulate the amplitude of tensor inhomogeneities
\begin{equation}
\rho_{A_{ij}} \equiv \frac{1}{16 \pi G} \tilde A_{ij} \tilde A^{ij}, \label{eqn:RhoAij}
\end{equation}
in analogy with the density $\rho$ for the matter content on the right hand side of the Hamiltonian constraint \eqn{eqn:HamCon}. This is equivalent to the energy density of the Isaacson energy momentum tensor \cite{Isaacson:1967zz} $t_{00} =  \frac{1}{32\pi G} \langle \dot{h}_{\mu \nu} \dot{h}^{\mu \nu} \rangle$, when the perturbations are small and one can average over one period in spacetime.\footnote{We gain two factors of 2 from the definition $-2 \alpha K_{ij} = \partial_t \gamma_{ij}$, and then a further $1/2$ from the averaging procedure (ie, the RMS factor).} However, our simulations are not in the perturbative limit, and the entire spacetime contains only one mode over which to average. Nevertheless, we found it to be a useful concept for keeping track of the level of perturbations, in practical aspects such as setting the AMR regridding conditions near collapse, and in our approximate model, which is described in section \ref{sec-interpretation}.

We note that there are two main energy scales, $H_0$ given by \eqn{eqn:Lformula}  which is the scale of inflation in the absence of any perturbations, and the actual Hubble scale $H_{act}^2 = (8/3)\pi G(\rho + \rho_{A_{ij}})$, where $\rho$ is given by \eqn{eqn:rhototal}. We will focus on small field inflation, and hence $\rho \approx V_0$, thus $H_{act}^2 \approx (8/3)\pi G(V_0 + \rho_{A_{ij}})$. As we will see, our simulations will scan through metric perturbations $\Delta A$ such that $\rho_{A_{ij}} \sim {\cal O}(\mathrm{few}) \times V_0$.  As a rule of thumb, $\Delta A=1 \times 10^{-10} \mpl$, gives an average for the metric fluctuations approximately equal to  $\langle \rho_{A_{ij}} \rangle \approx V(\phi)$. We represent the this scanning parameter as the absolute ratio $\rho_{A_{ij}}/V_0$.

\section{Results} \label{sect:results}

In this section we summarise our results and propose an explanation for the effects observed.

\subsection{Key findings - small field}

Our main result is shown in figure \ref{fig:Efolds2D}.  Adding the tensor background as described above results in fewer $e$-folds than in the case in which the inhomogeneities were confined to the inflaton field and the conformal factor. As we mentioned in the introduction, inflations fail when the scalar field falls out of the inflationary plateau into the reheating minimum across the entire space, thus ending the inflationary dynamics. 

However, interestingly, a non-zero fluctuation in the scalar field ``seed'' was required for early failure - simply adding a tensor background to a spatially homogeneous field was insufficient to cause failure even at large values of $\Delta A$.  While in this case inhomogeneities in the field did develop, they were extremely small and developed slowly, by which time slow-roll was well established. This is because the Klein-Gordon equation for the scalar field couples to the metric only through the $\gamma_{ij}$ and $K$ terms. 

\begin{figure}
\begin{center}
\includegraphics[width=8cm]{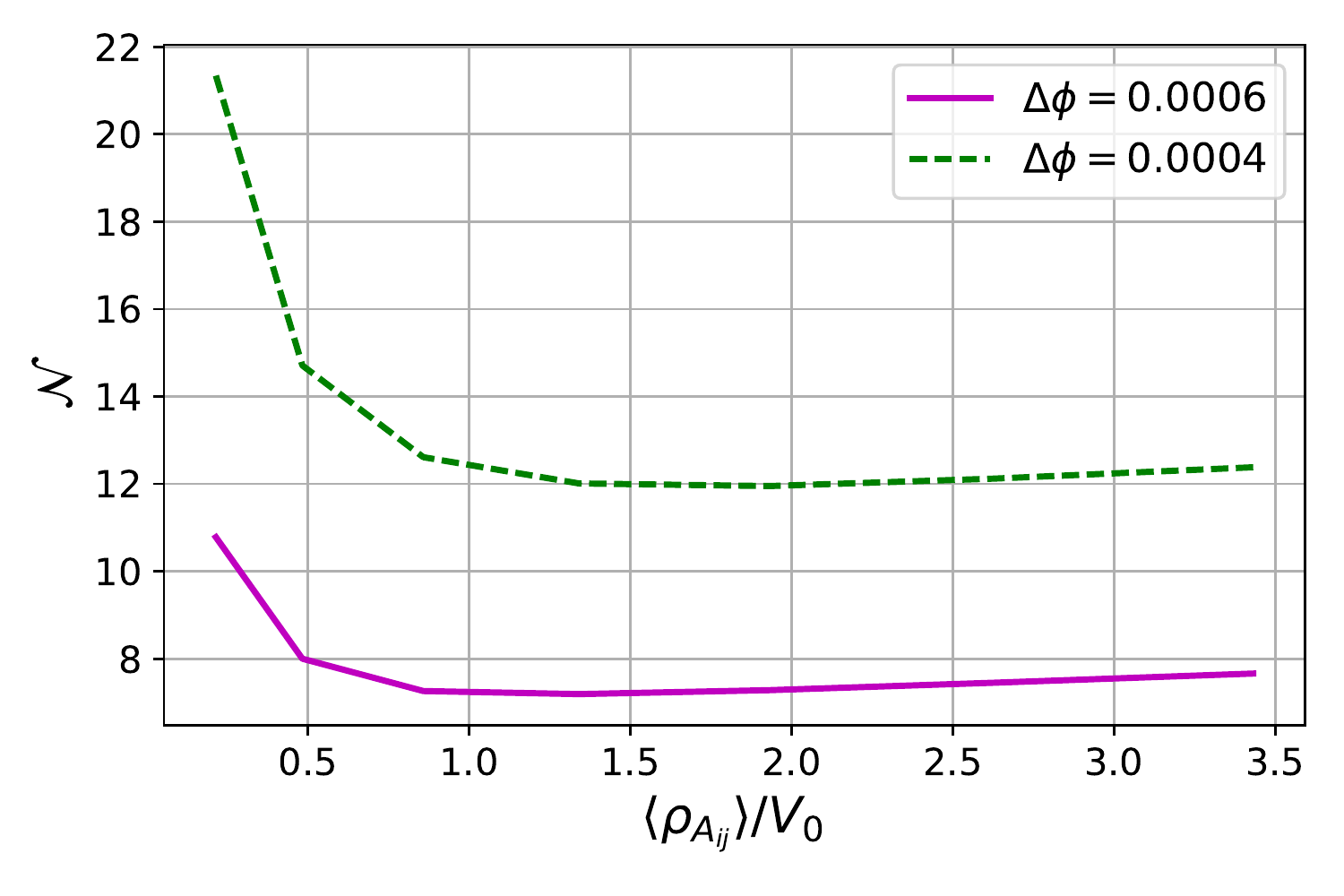}
\caption{Number of $e$-folds at the failure point (where inflation first ends), versus the amplitude of the fluctuations in $\bar A_{ij}$, for different values of $\Delta \phi$ in the initial inflaton field inhomogeneities (for $n=1$). It can be seen that the initially downward trend levels off and even reverses slightly as the amplitude is increased.
\label{fig:Efolds1D}}
\end{center}
\end{figure}

For $\langle\rho_{A_{ij}}\rangle\gtrsim V(\phi_0)$ the reduction in $e$-folds levels out, and increases in the extrinsic curvature fluctuations do not reduce inflation further. The downward trend even reverses as the amplitude of the fluctuations increases, which can be seen more clearly in figure \ref{fig:Efolds1D}.

Above  $\langle \rho_{A_{ij}} \rangle/V_0 \sim 4$ the perturbations reach a critical level above which regions of spacetime undergo collapse. The collapsing regions were roughly planar and difficult to evolve until the end of inflation, possibly due to the formation of coordinate singularities and other gauge issues.\footnote{It is also not clear whether the $1+log$ slicing condition used should be able to stably evolve such collapses - see \cite{Hilditch:2013cba} for simulations in this direction.} However, as long as we can evolve them, they do not appear to disrupt the inflation in surrounding areas. Intuitively, we expect the collapse to eventually lead to formation of black holes which would ``inflate away'' (as was seen in the large field case of \cite{Clough:2016ymm}), although we did not evolve them to their end states. To some extent the difficulty arises because the initial conditions for the inhomogeneities are too ``regular''. For a more random initial configuration (ie, a superposition of many modes with random phases), we would expect that the collapsing regions should be less planar and more contained in roughly spherical regions, which should make simulating the full collapse easier. We will investigate more general inhomogeneities in future work.

\begin{figure}
\begin{center}
\includegraphics[width=8cm]{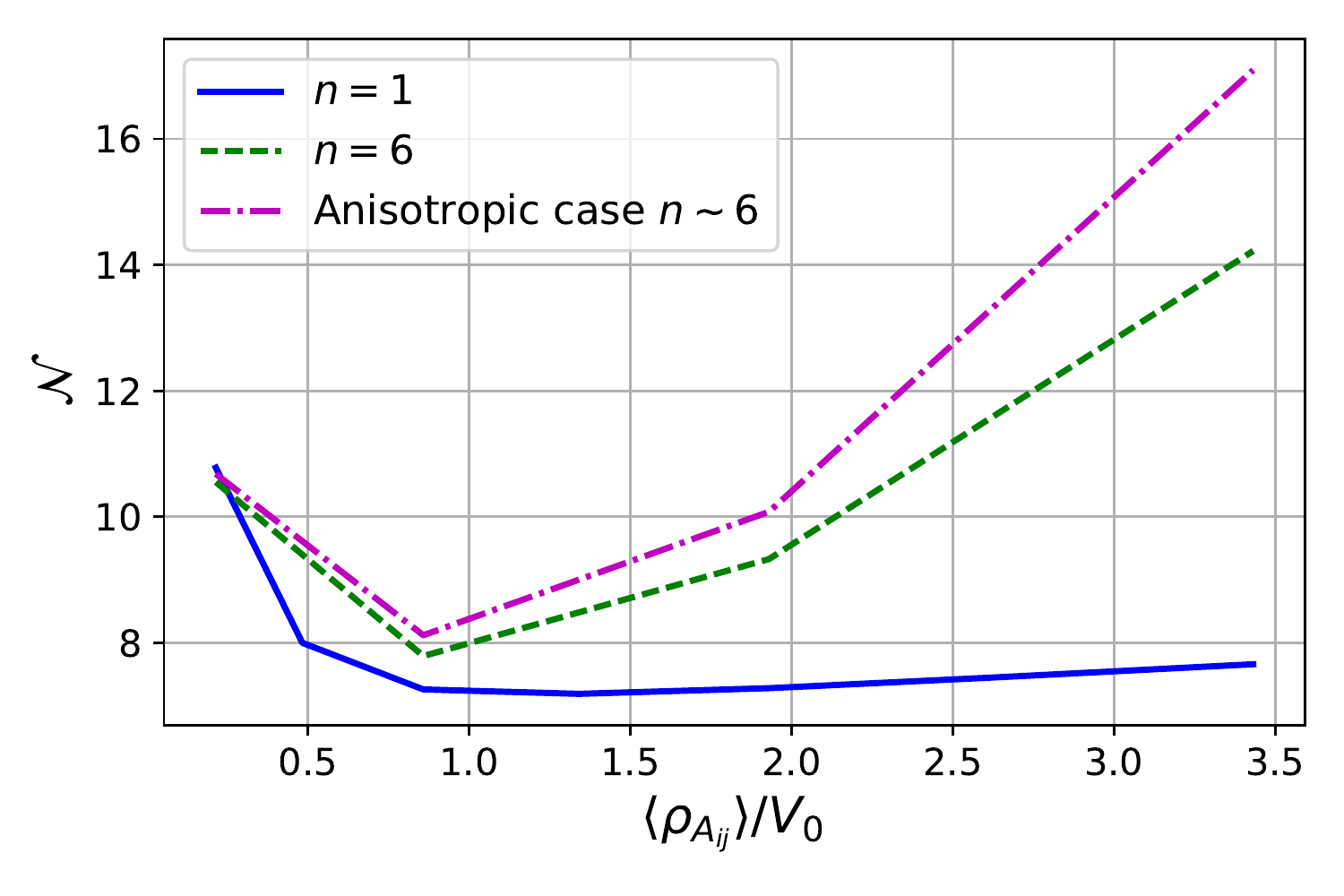}
\caption{Number of $e$-folds at the failure point (where inflation first ends), versus the amplitude of $\langle \rho_{A_{ij}} \rangle/V_0$, for different values of $n$ in the initial inflaton field inhomogeneities (for $\Delta \phi=0.0006 \mpl$). We compare the cases for tensor perturbations for $n=6$, $n=1$, and the anisotropic superposition of modes with $n\sim6$. It can be seen that the overall behaviour and reduction in values is similar, although the recovery for higher $\Delta A$ is stronger for the isotropic $n=6$ case, and slightly higher again for the anisotropic case.
\label{fig:Nefolds_nis6}}
\end{center}
\end{figure}

In the case of fluctuations in $\bar A_{ij}$ with $n=6$, we saw no significant qualitative difference in the results for the number of $e$-folds, although the recovery with increasing $\Delta A$ was somewhat stronger, as we can see in figure \ref{fig:Nefolds_nis6}. This figure also shows the behaviour for the superposition of modes, which is very similar to the simpler $n=6$ case. This shows that more anisotropic conditions do not give a significantly different behaviour, justifying our use of the simpler model of fluctuations in the rest of the work.

Since the behaviour in each case is broadly similar, this implies that the particular length scale and anisotropy of the fluctuations does not have a strong effect - it is the average initial energy density that has the strongest role to play in reducing the number of $e$-folds. We will use this observation in constructing a simple numerical model, in section \ref{sec-interpretation}.

\begin{figure}
\begin{center}
\includegraphics[width=8cm]{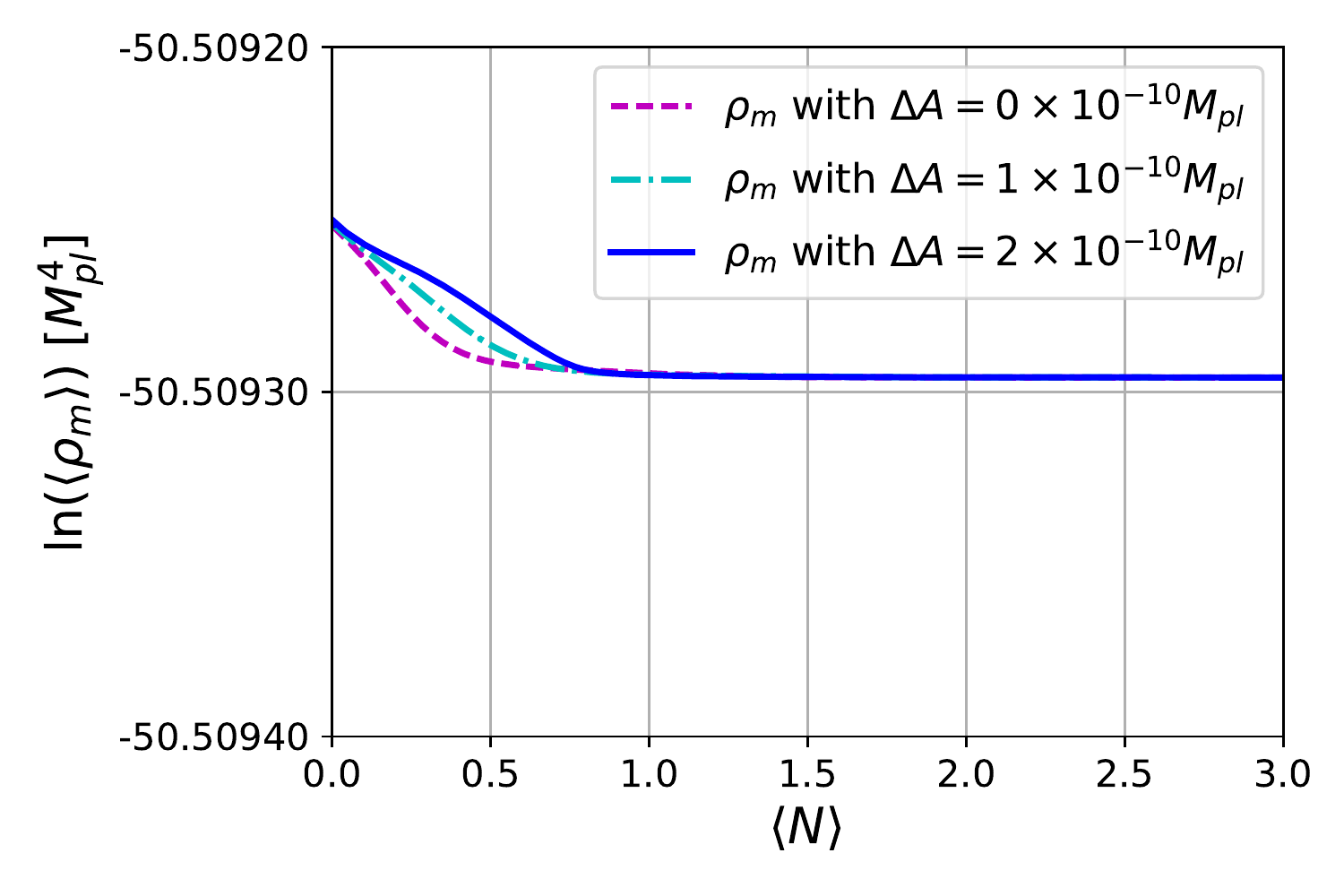}
\caption{Evolution of the (matter) energy density versus the average number of $e$-folds across the grid, in the cases of different initial amplitudes $\Delta A$. One can see that the evoution of the matter energy density is only weakly affected by the presence of the gravitational wave fluctuations, and still settles into the constant inflationary period within one e-fold.
\label{fig:SFrho_vs_N}}
\end{center}
\end{figure}

\begin{figure}
\begin{center}
\includegraphics[width=8cm]{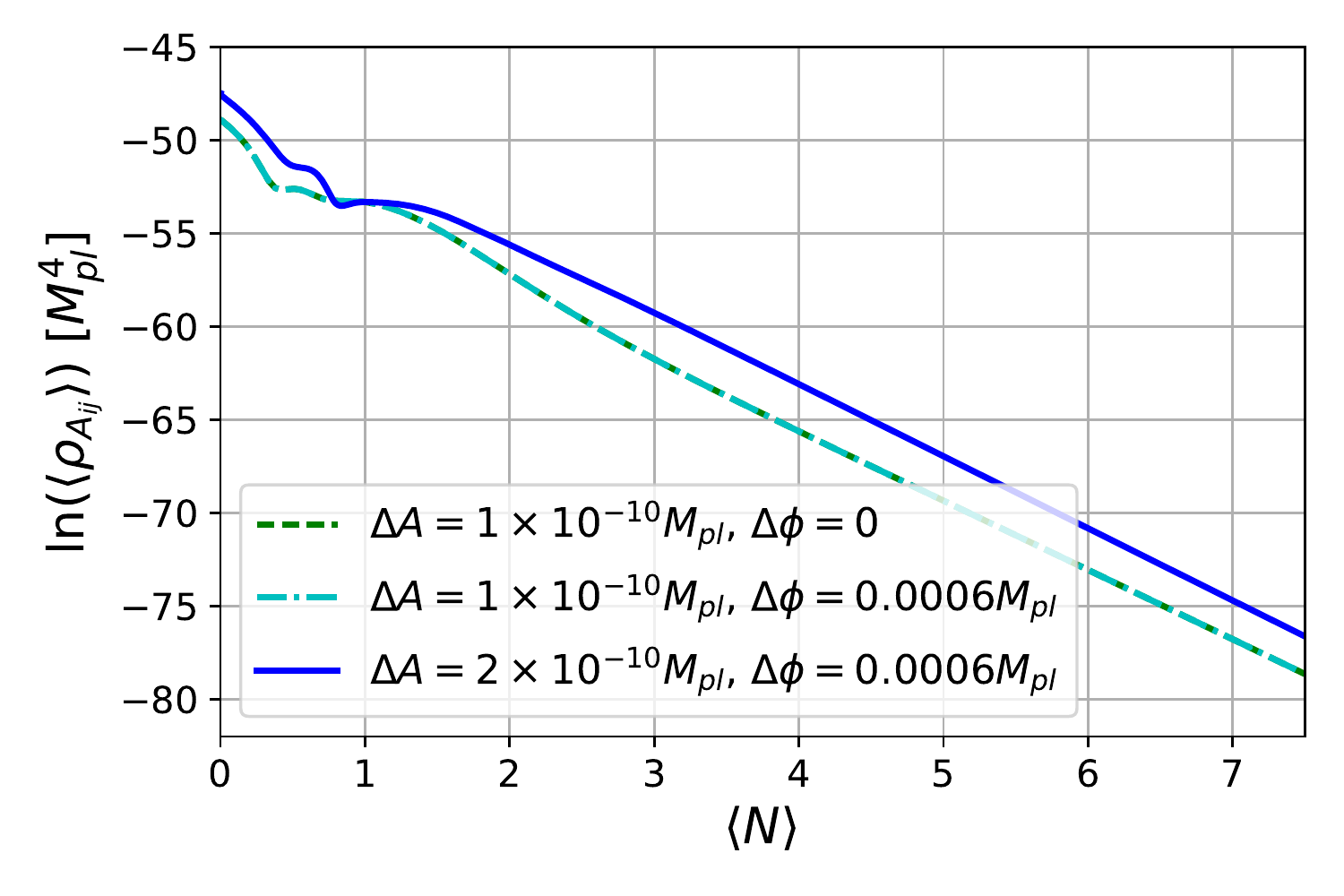}
\caption{Evolution of the average tensor perturbation energy density $\langle \rho_{A_{ij}} \rangle$ versus the average number of $e$-folds across the grid $\langle \mathcal{N} \rangle$, in the cases of different fluctuations in $\bar A_{ij}$ and $\phi$, with horizon sized modes in both (ie, $n=1$). The behaviour is unaffected by the gradient energy of the inflaton field - the lines for $\Delta \phi=0$ and $\Delta \phi=0.0006 \mpl$ with the same value of $\Delta A$ are indistinguishable.
\label{fig:SFrhoAij_vs_N}}
\end{center}
\end{figure}

The resulting evolution of the average matter energy density $\langle \rho_m \rangle$, and the metric energy density $\langle \rho_{A_{ij}} \rangle$ are shown in figures \ref{fig:SFrho_vs_N} and \ref{fig:SFrhoAij_vs_N} respectively. We can see that the matter evolution is somewhat affected by the gravitational wave background but not vice versa - the evolution of $\langle \rho_{A_{ij}} \rangle$ does not appear to depend on the perturbations in $\phi$, which is not surprising as the scalar field gradient energy density is still strongly subdominant to $V(\phi)$. 

At early times, $\langle \rho_{A_{ij}}\rangle$ behaves irregularly though overall it roughly scales as  $a^{-6}$, while at later times, once de Sitter expansion has begun, $\langle \rho_{A_{ij}} \rangle \propto a^{-4}$ consistent with radiation. This is due to the fact that modes of $A_{ij}$ have wavelengths comparable to Hubble, and do not support a wave-like solution.

\begin{figure}
\begin{center}
\includegraphics[width=8cm]{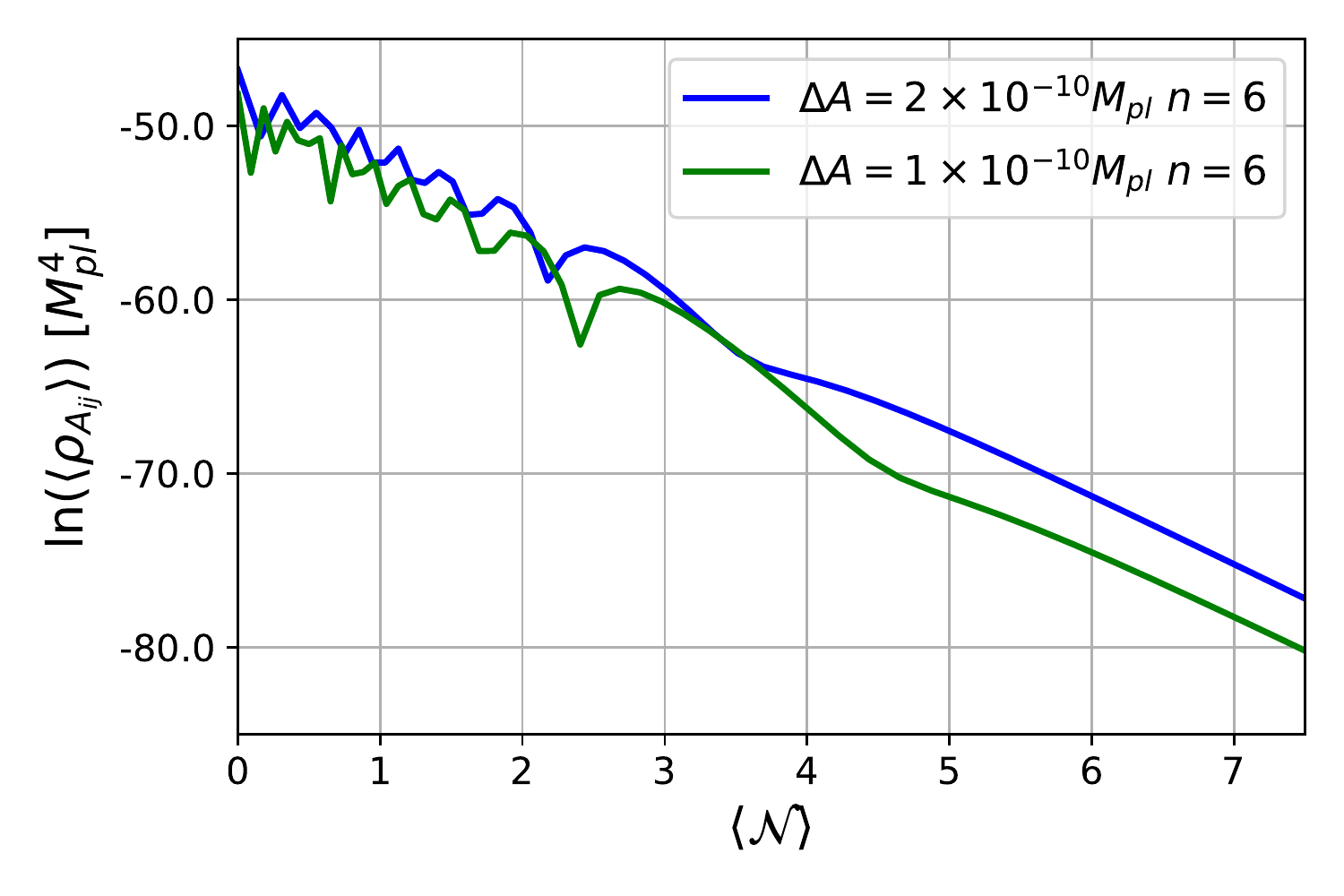}
\caption{Evolution of the average tensor perturbation energy density $\langle \rho_{A_{ij}} \rangle$ versus the average number of $e$-folds across the grid $\langle \mathcal{N} \rangle$, for $n=1$ modes in the scalar field and smaller wavelength, $n=6$ modes for the metric perturbations. The evolution of $\langle \rho_{A_{ij}} \rangle$ is again complex in the early stages as in the case of $n=1$, now showing strong oscillations in the average value, but still scales as $a^{-4}$ overall.
\label{fig:SFrhoAij_vs_N_nis6}}
\end{center}
\end{figure}

The same plot is given for $n=6$, where we would perhaps expect a more homogeneous evolution. We see in figure \ref{fig:SFrhoAij_vs_N_nis6} that the evolution of $\langle \rho_{A_{ij}} \rangle/V_0$ is again complex in the early stages - now showing strong oscillations in the average value, but scales as $a^{-4}$ overall as one would expect for subhorizon modes.


If $\langle \rho_{A_{ij}} \rangle > V_0$ in the early stages, spacetime is either dominated by scalar kinetic energy ($\rho \propto a^{-6}$) or radiation dominated ($\rho \propto a^{-4}$), and one might suspect that the inflaton -- instead of slowly rolling -- be attracted to the dominant dynamic and roll coherently down the potential resulting in the loss of total $e$-folds. However, it is easy to check that this is not the case -- in the limit when $\langle \rho_{A_{ij}} \rangle \gg V_0$ and $\langle \rho_{A_{ij}} \rangle = \rho_A^0a^{-m}$, the homogeneous Klein-Gordon equation behaves as 
\begin{equation}
\phi'' + \left(4-\frac{m}{2}\right){\phi'}{a} - 3a^{m-2}\sqrt{\epsilon}\frac{V_0}{\rho_A^0} = 0 \label{eqn:kinationdom}
\end{equation}
where primes denote $d/da$, $\epsilon$ is the slow roll parameter and we have assumed that the scalar field rolls in the positive direction down the potential). The middle term acts as a friction, while the last term is proportional to $V_0/\rho_A^0$, i.e. the scalar field rolls even less when dominated by the metric energy. Solving \eqn{eqn:kinationdom} numerically shows that the early metric energy domination does not change the coherent scalar dynamics significantly.

\begin{figure}
\begin{center}
\includegraphics[width=8cm]{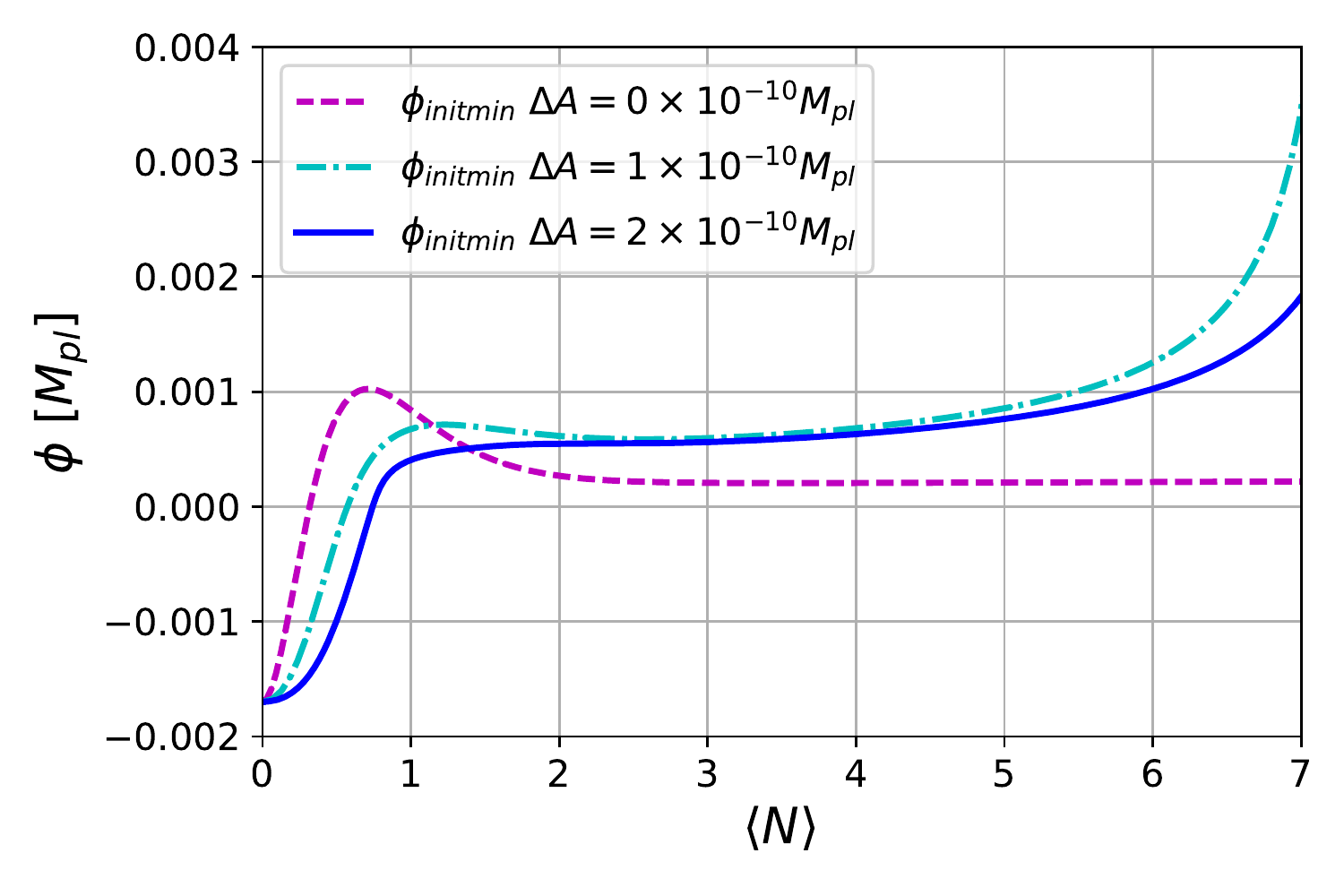}
\caption{Evolution of the initial minimum point of $\phi$ in the cases of different initial fluctuations in $A_{ij}$. We can see that as the fluctuations $\Delta A$ increase, the pullback of the field is reduced.
\label{fig:SFphi_vs_N}}
\end{center}
\end{figure}

\begin{figure}
\begin{center}
\includegraphics[width=8cm]{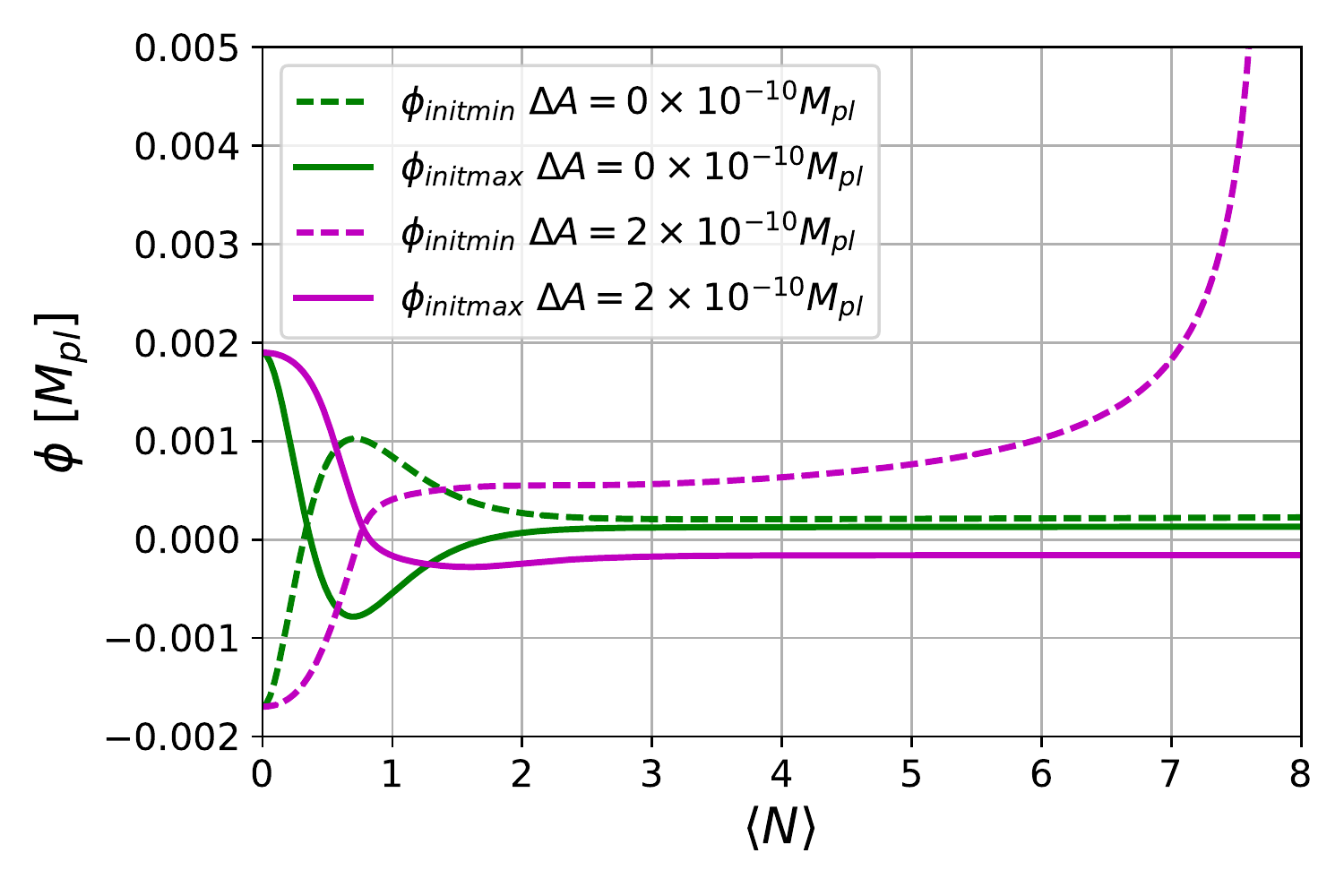}
\caption{Evolution of the initial maximum and minimum points of $\phi$ in the cases of different initial fluctuations in $A_{ij}$. We can see that as the fluctuations $\Delta A$ increase, the pullback of the field to a flat configuration is reduced, and there is a more rapid failure at later times.
\label{fig:SFphiMM_vs_N}}
\end{center}
\end{figure}

Rather, to understand the loss of $e$-folds, it is most instructive to look at the evolution of the failure point (the point at which inflation first ends, which initially has the minimum value of $\phi$). This is shown in figure \ref{fig:SFphi_vs_N}. The failure mode in which the field falls off the potential hill, happens quicker than in the absence of the GW content, and there is a reduction in the initial pullback on the field. One can also look at both the initial maximum and minimum points of the field in figure \ref{fig:SFphiMM_vs_N}. This behaviour helps us to explain the scaling of the $e$-folds with increasing $\Delta A$, given the model we propose in section \ref{sec-interpretation}.

\subsection{Key findings - large field}

We also performed several simulations in the large field inflation case, with an $m^2 \phi^2$ potential as described above, to confirm that additional metric perturbations did not interfere with large field inflation.

As expected, inflation still begins, and lasts the maximum number of $e$-folds in the majority of the spacetime, with only locally collapsing regions of high density. Thus, the behaviour was not strongly affected by the presence of the tensor modes.

\begin{figure}
\begin{center}
\includegraphics[width=7.8cm]{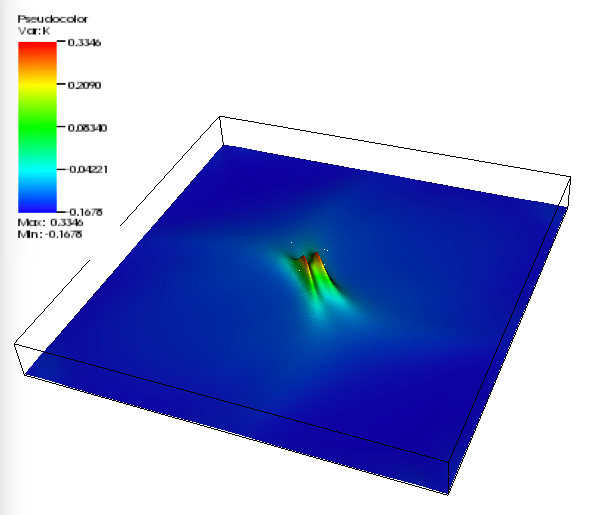}
\includegraphics[width=7.8cm]{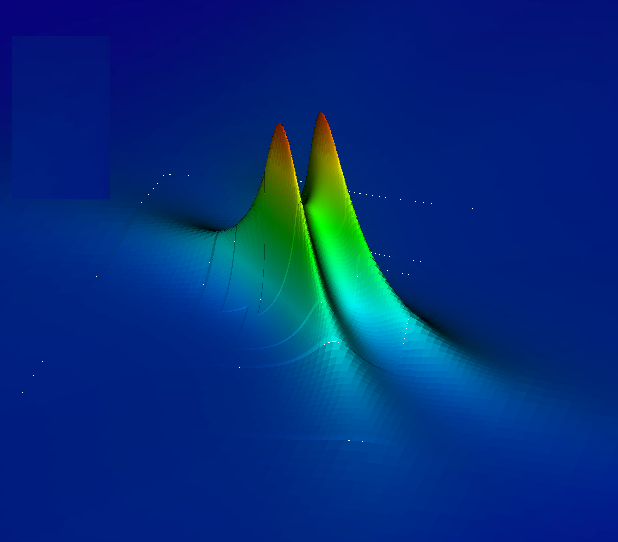}
\includegraphics[width=7.8cm]{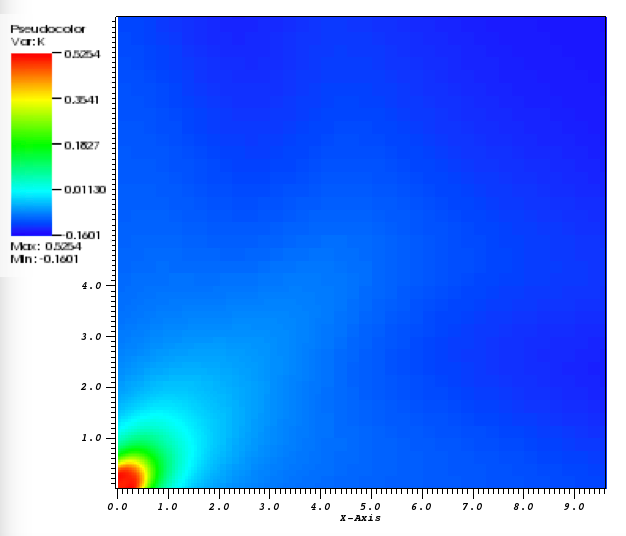}
\caption{Large field simulations showing the value of the extrinsic curvature $K$. The red regions are collapsing and the blue regions are inflating. We still see collapse in the centre and corner regions, as in the absence of tensor perturbations, but these regions are less symmetric, and the black holes (identified using an apparent horizon finder) tended to form more readily in the corner regions. Overall the qualitative behaviour of local collapses in a globally inflating spacetime is unchanged.
\label{fig:LF_K}}
\end{center}
\end{figure}

One difference, as illustrated in figure \ref{fig:LF_K}, was the formation of black holes. In the absence of the tensor modes, we would form two identical black holes - one in the center of the grid and one in the corners (the grid is periodic, so the corners represent the same black hole). With the presence of tensor perturbations the symmetry was broken, and while both regions still collapsed to some extent, the corners now formed black holes more readily.

\subsection{Interpretation of results}
\label{sec-interpretation}

In our previous paper \cite{Clough:2016ymm} we derived a limit on the size of $\Delta \phi$ above which inflation would immediately fail, as a function of the features of the inflationary potential and the size of the modes. Below this limit the field tended to oscillate into a flat configuration, from which inflation could proceed. The critical amplitude was based on equating the pullback force from the initial field gradients to the potential gradient, which tended to pull the field down into the reheating minimum, and thus it should remain unchanged by the presence of a background of tensor perturbations.

However, as we have seen in this work, \emph{below} this critical limit for $\Delta \phi$, the picture changes with the addition of the tensor perturbations. The oscillations of the maximal points in the field get stuck at values further from the flat configuration. These points do not initially result in failure and may still enter a slow roll period, but they accelerate the failure at the critical point at later times. As we continue to increase the amplitude of the metric perturbations, things actually improve again with the field stuck closer to the flat configuration. 

To confirm this picture, we again consider the evolution of the extremal points. The Klein-Gordon equation is
\begin{equation}
\ddot{\phi} + 3H \dot{\phi} - \gamma^{ij}\partial_i \partial_j\phi = - V' ~,
\end{equation}
where we have included the gradient term in addition to the usual friction term\footnote{Compared to our discussion of the pullback effect in \cite{Clough:2016ymm}, we have included the friction term.}.

We make the simplifying assumption that \emph{the gravitational wave background can be modelled by a homogeneous energy density}, equal to the initial homogeneous average across the spatial slice, and scales like 
\begin{equation}
\langle \rho_{A_{ij}} \rangle/V_0 = r  ^{-p} ~,
\end{equation}
where $r$ sets the initial value of the energy density relative to the inflationary energy density $V_0$, and $p \sim 4$ is the power law scaling of the field with the expansion. We also assume that the initial spacetime is conformally flat, and that  $\chi$ is approximately constant around the critical point, so that it can be represented by some spatially homogeneous number of $e$-folds, $a$, with $\chi = 1/a^2$.

It is helpful to rewrite the Klein Gordon equation with the derivatives taken with respect to the number of $e$-folds. Defining the deviation from the average value at the critical failure point as $x \equiv \phi - \phi_0$, and assuming $\dot\phi_0$ is small, then the Klein Gordon equation
becomes, for our configuration with single horizon scale modes,
\begin{equation}
\ddot{x} + 3H \dot x + \frac{4 \pi^2 H_0^2}{a^2} x = - V' \label{eqn:HarmonicOscillator}
\end{equation}
which can be written with ``time'' instead measured by the number of $e$-folds $N$ as
\begin{equation}
H^2 \frac{d^2x}{dN^2} + A \frac{dx}{dN} + B x = - V', \label{eqn-dxdN}\,.
\end{equation}
Here the values of the coefficients A and B are
\begin{equation}
A  = 3H_0^2 \left[ 1 + r e^{-pN} \left(1 - \frac{p}{6} \right) \right] \label{eqn-dxdN_A}
\end{equation}
and
\begin{equation}
B  = \frac{4 \pi^2 H_0^2}{e^{2N}} ~. \label{eqn-dxdN_B}
\end{equation}
In analogy with a simple damped harmonic oscillator, the value of $A$ represents the (Hubble) friction, and the value of $B$ the restoring pullback force, while the potential gradient provides the driving force which pushes the field down the hill. It is clear that for $r>0$ the friction is increased at early times, as well as the ``mass'' $H_0^2$, which for large $r$ reduces the ``overshoot'' in the initial transient behaviour. At later times the values are consistent with the case of $r=0$, meaning that if the field survives the inital period, it can still settle into slow roll.

\begin{figure}
\begin{center}
\includegraphics[width=8cm]{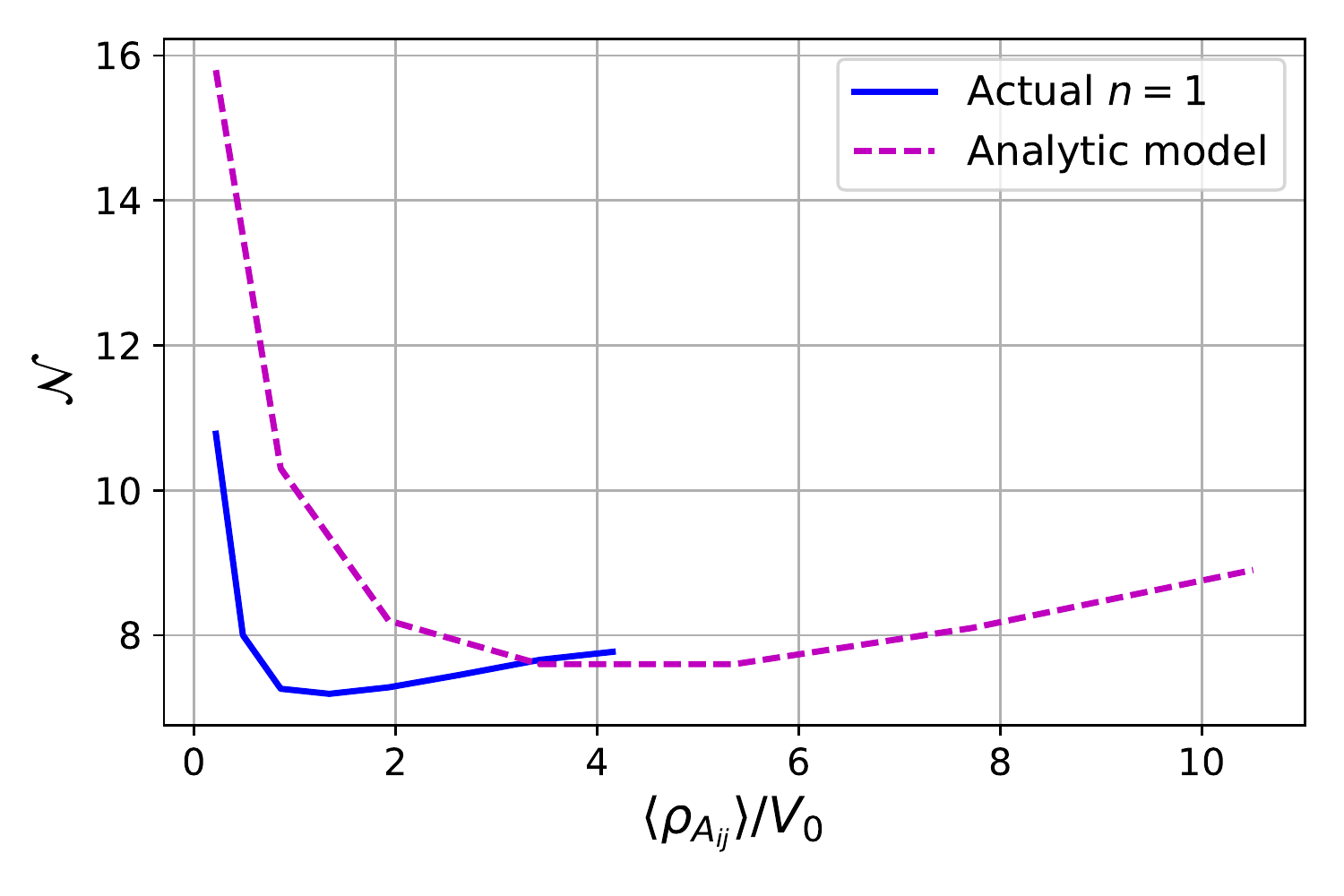}
\caption{Number of $e$-folds at the failure point (where inflation first ends), versus the amplitude of the fluctuations in $\bar A_{ij}$. We compare the case of tensor perturbations with $n=1$ (horizon sized modes) to the simple numerical model described in section \ref{sec-interpretation} with a $p=4$ power law scaling for the average energy density $\langle \rho_{A_{ij}} \rangle$. We see that the model recreates the qualitative shape of the curve.
\label{fig:Nefolds_analytic}}
\end{center}
\end{figure}

\begin{figure}
\begin{center}
\includegraphics[width=8cm]{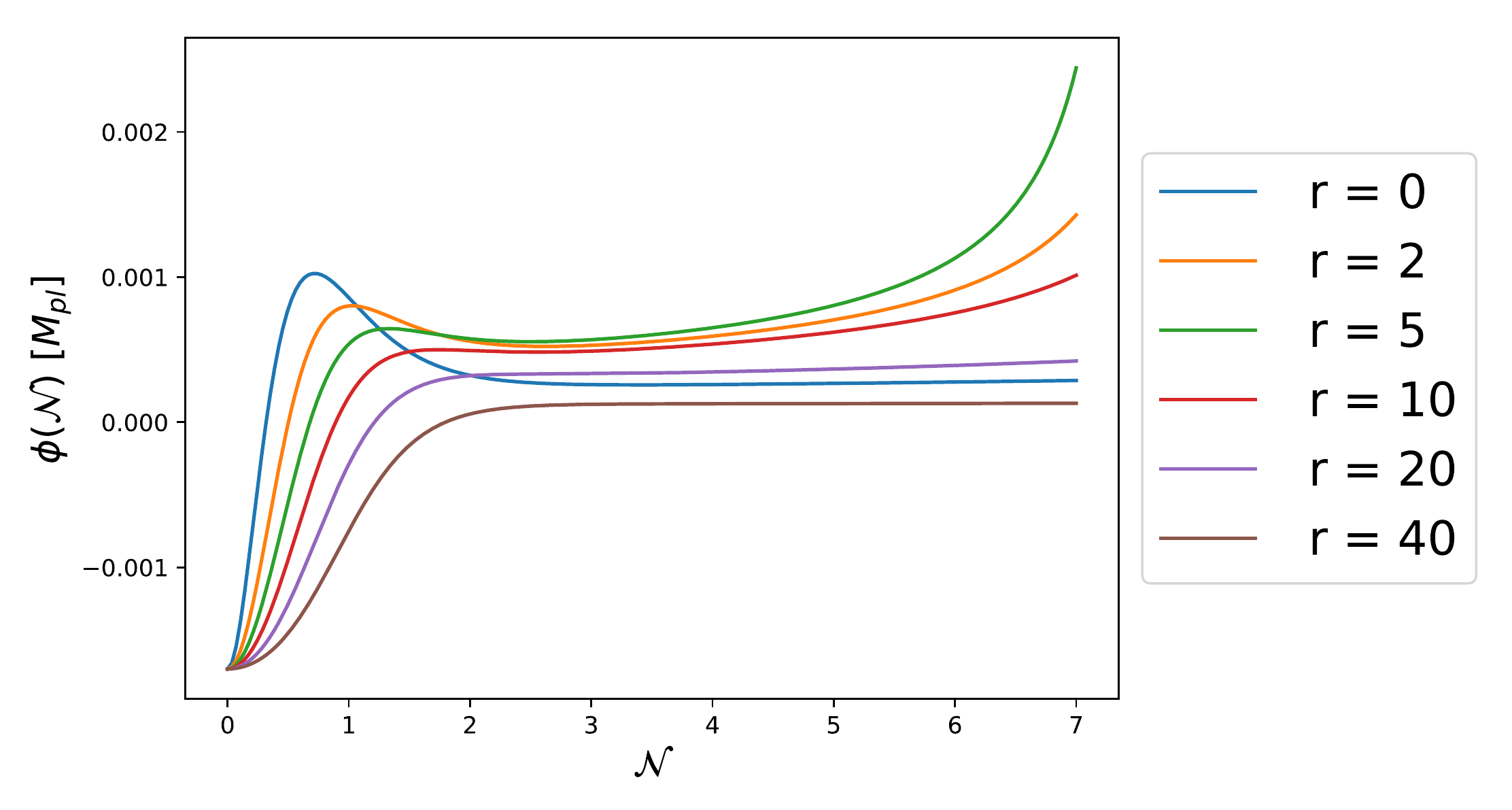}
\caption{Evolution of the initial minimum point of $\phi$ in the cases of different initial fluctuations in $A_{ij}$, paraameterised by $r=\langle \rho_{A_{ij}} \rangle/V_0$, using the simple numerical model with $p=4$. Comparing this with figure \ref{fig:SFphi_vs_N}, we can see that the qualitative behaviour of the full simulations is well reproduced by the simple model.
\label{fig:AnalyticR}}
\end{center}
\end{figure}

\eqn{eqn-dxdN} may be solved very easily using numerical integration. The results are shown in figure \ref{fig:Nefolds_analytic} for the number of $e$-folds at failure (the point at which $\phi=\phi_*$) versus the actual results obtained for the case of a single mode $n=1$. The evolution of the field is shown in figure \ref{fig:AnalyticR}. It is clear that the results are qualitatively the same as the full evolution, and thus this fairly simple model can indeed be used to gain intuition about the dynamics of the system in the presence of inhomogeneous tensor modes.

Note that while $\rho_{A_{ij}}$ can vary significantly at different locations, our simple model is qualitatively accurate, which suggests that the main driver of the behaviour is the change in the average Hubble expansion rate, rather than the detailed dynamics of the tensor perturbations. 

\begin{figure}
\begin{center}
\includegraphics[width=8cm]{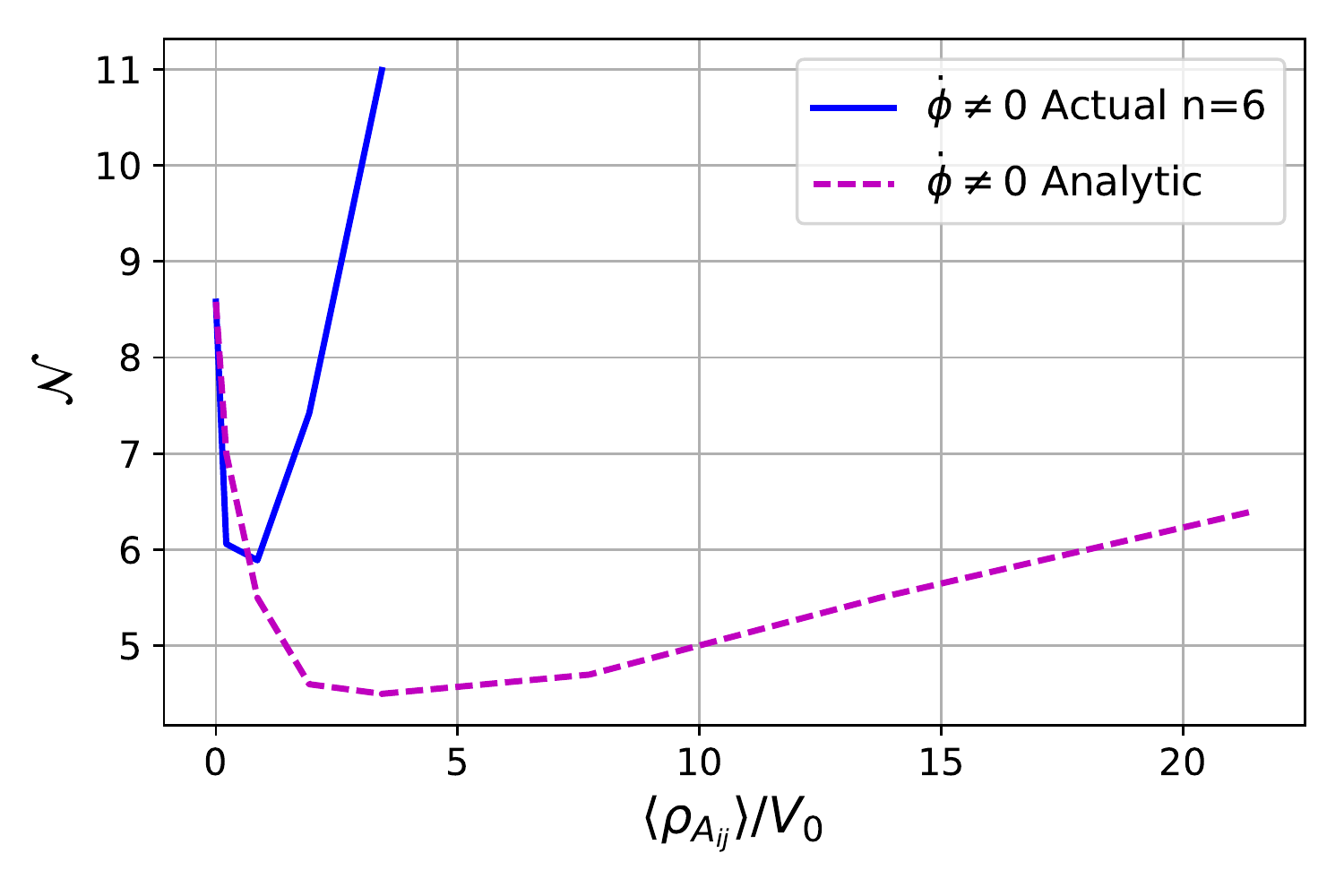}
\caption{Number of $e$-folds at the failure point (where inflation first ends), versus the amplitude of the fluctuations in $\bar A_{ij}$, with non zero $\dot\phi$. The simulation used the $n=6$ case in which a stronger recovery in the number of $e$-folds was seen than for $n=1$. The simple numerical model is that described in section \ref{sec-interpretation} with a $p=4$ power law scaling for the average energy density $\langle \rho_{A_{ij}} \rangle$, where we chose an initial value for $\dot\phi$ which gave the same number of $e$-folds in the $\rho_{A_{ij}}=0$ case as the actual simulations. We see the same trends in the analytic and actual results, although as before the model appears to be more conservative in estimating the number of $e$-folds.
\label{fig:Nefolds_analytic2}}
\end{center}
\end{figure}

This observation raises the interesting possibility that a gravitational wave background may be able to mitigate the effect of an initial non zero value of $\dot\phi$. In general, the addition of even a small velocity of the field down the potential will disrupt slow roll for small field models. By increasing the value of $H$ initially using tensor perturbations, perhaps we can reduce this effect. Figure \ref{fig:Nefolds_analytic2} shows the results of simulations to test this possibility. We see that we can indeed mitigate the effect of a non zero $\dot\phi$, with a tensor background increasing the number of efolds from 8.6 in their absence to 11.0 with $\Delta A=0.2$ amplitude tensor perturbations present. As in our earlier simulations we were limited in how large we could make our fluctuations. However, our analytic model allows us to probe the behaviour further, and we see that it implies that we can continue to increase the number of e-folds in cases with non zero $\dot\phi$, provided the field does not immediately fail.

One additional approximation in our model is useful to elucidate the findings referred to in our previous work regarding convexity of the potential. If we expand the potential gradient as a Taylor series about the average value $\phi_0$
\begin{equation}
V' = V '\rvert_{\phi=\phi_0} + V'' \rvert_{\phi=\phi_0} x + O(x^2)~, \label{eqn-dVdphi}
\end{equation}
then equation \ref{eqn-dxdN} becomes
\begin{equation}
H^2 \frac{d^2x}{dN^2} + A \frac{dx}{dN} + (B + V'' \rvert_{\phi=\phi_0}) x = - V' \rvert_{\phi_0}, \label{eqn-dxdN2}
\end{equation}
This makes clear that a \emph{convex} potential $V'' > 0$ serves to \emph{increase} the restoring force, making the model more robust to inhomogeneities, while a concave one will \emph{reduce} the restoring force, and so tend to destabilise the slow roll. Thus a potentially useful distinction of models of inflation when discussing stability is into convex and concave potentials. 

\section{Conclusions} \label{sect:conclusions}

In general, we find a somewhat complicated picture on the effect of tensor perturbations. Unlike for perturbations in the scalar inflaton field, increased tensor perturbations do not necessarily lead to a decrease in the number of e-folds of inflation, and we see that they can in fact mitigate other problematic initial conditions for slow roll such as having a non zero value of $\dot \phi$.

As in our previous paper, we find that the scalar field dynamics are the key driver of the behaviour in small field/concave models, and that our more complex simulations can be approximated by relatively intuitive toy models. 

Specific results were summarised in the Introduction, section \ref{sec-intro}, and illustrated in figure \ref{fig:Efolds2D}.

It would be interesting in future to consider more general initial conditions with superpositions of modes with different wavelengths, as there are hints in our results that this may improve the recovery of slow roll when increasing the amplitude of tensor fluctuations. More random initial conditions would also allow us to probe further the collapses which we observe at higher amplitudes, to confirm that they result in black holes which are then inflated out of the spacetime. In addition, it would be interesting to refine the toy models to gain a better analytic understanding.

\vspace{1cm}

\acknowledgments

We thank Jens Niemeyer, Helvi Witek, Pedro Ferreira, Josu Aurrekoetxea and Thomas Helfer for helpful conversations. EAL acknowledges support from an STFC AGP grant ST/ P000606/1. RF is supported in part by the Alfred P. Sloan Foundation, the Department of Energy under Grant No. DE-SC0009919, and a grant from the Simons Foundation/SFARI 560536.

We would also like to thank the GRChombo team (http://grchombo.org/collaborators.html), in particular Markus Kunesch, for their work on the code, and the COSMOS team at DAMTP, Cambridge for their ongoing technical and developmental support in collaboration with Intel. 

The authors gratefully acknowledge the computer resources at Marenostrum IV, Finis Terrae II and LaPalma and the technical support provided by the Barcelona Supercomputing Center via the PRACE grant Tier-0 PPFPWG, by the Supercomputing Centre of Galicia and La Palma Astrophysics Centre via BSC/RES grants AECT-2017-2-0011 and AECT-2017-3-0009. Simulations were also performed on the COSMOS supercomputer, part of the DiRAC HPC, a facility which is funded by STFC and BIS, the GWDG cluster in G{\"o}ttingen. Our visualisations use the VisIt \cite{HPV:VisIt} and yt \cite{yt2011} software packages.


\appendix

\section{GRChombo code}
\label{sect:AppendixA}

This appendix summarises the key features of the numerical relativity code $\textsc{GRChombo}$. For a more full discussion see \cite{Clough:2015sqa}, and the $\textsc{GRChombo}$ website at \url{http://grchombo.org}, which includes links to movies of simulations using the code.

\subsection{Numerical implementation}
$\textsc{GRChombo}$ is a multi-purpose numerical relativity code, which is built on top of the open source $\mathtt{Chombo}$ framework. $\mathtt{Chombo}$ is a set of tools developed by Lawrence Berkeley National Laboratory for implementing block-structured AMR in order to solve partial differential equations \cite{Chombo}.

The key features of $\mathtt{Chombo}$  are:
\begin{itemize}
\item{\emph{C++ class structure}: $\mathtt{Chombo}$ is primarily written in the C++ language, using the class structure inherent in that language to separate the various evolution and update processes.}
\item{\emph{Adaptive Mesh Refinement}: $\mathtt{Chombo}$ provides Berger-Oliger style \cite{bergeroliger,BergerColella} AMR with Berger-Rigoutsos \cite{BergerRigoutsis91} block-structured grid generation. Chombo supports full non-trivial mesh topology -- i.e. many-boxes-in-many-boxes. The user is required to specify regridding criteria, which is usually based on setting a maximum threshold for the change in a variable across a gridpoint.}
\item{\emph{MPI scalability}: $\mathtt{Chombo}$ contains parallel infrastructure which gives it the ability to scale efficiently to several thousand CPU-cores per run. It uses an inbuilt load balancing algorithm, with Morton ordering to map grid responsibility to neighbouring processors in order to optimize processor number scaling.}
\item{\emph{Standardized Output and Visualization}: $\mathtt{Chombo}$ uses the $\mathtt{HDF5}$ output format, which is supported by many popular visualization tools such as $\mathtt{VisIt}$. In addition, the output files can be used as input files if one chooses to continue a previously stopped run -- i.e. the output files are also checkpoint files.}
\end{itemize}

The key features of $\textsc{GRChombo}$ are:
\begin{itemize}
\item{\emph{BSSN formalism with moving puncture}: $\textsc{GRChombo}$ evolves the Einstein equation in the BSSN formalism with scalar matter. Singularities of black holes are managed using the moving puncture gauge conditions \cite{Campanelli:2005dd, Baker:2005vv}. These evolution equations and gauge conditions are detailed further below. There is an option to turn on CCZ4 constraint damping terms if required, but this was not used in this work.}
\item{\emph{4th order discretisation in space and time}: We use the method of lines with 4th order spatial stencils and a 4th order Runge-Kutta time update. We use symmetric stencils for spatial derivatives, except for the advection derivatives (of the form $\beta^i \partial_i F$) for which we use one-sided/upwinded stencils. In \cite{Clough:2015sqa} it was shown that the convergence is approximately 4th order without regridding, but reduces to 3rd order convergence with regridding effects.}
\item{\emph{Kreiss-Oliger dissipation}: Kreiss-Oliger dissipation is used to control errors, from both truncation and the interpolation associated with regridding.}
\item{\emph{Boundary conditions}: We use either periodic boundaries or Sommerfeld boundary conditions \cite{Alcubierre:2002kk}, which allow outgoing waves to exit the grid with minimal reflections. For many simulations, the AMR ability allows us to set the boundaries far enough away so that reflections do not affect the results during simulation time. In this work only periodic boundary conditions were used.} 
\item{\emph{Initial Conditions}: In principle any initial conditions can be used, for example, where solutions to the constraints have been found numerically, these can be read into the grid using a simple first order interpolation. Note that $\textsc{GRChombo}$ itself does not currently solve the constraints for the initial conditions, although it can be used to relax the Hamiltonian constraint for the value of the conformal factor $\chi$ where the other variables are assumed to solve the momentum constraint and admit solutions consistent with the boundary conditions.}
\item{\emph{Diagnostics}: $\textsc{GRChombo}$ permits the user to monitor the Hamiltonian and momentum constraint violation, find spherically symmetric apparent horizons, extract gravitational waves and calculate ADM mass and momenta values.}
\end{itemize}

\subsection{Gauge choice}

$\textsc{GRChombo}$ uses the BSSN formalism \cite{Baumgarte:1998te,Nakamura:1987zz,Shibata:1995we} of the Einstein equation in 3+1 dimensions. This is similar to the more well known ADM decomposition \cite{PhysRev.116.1322}, but is more stable numerically. The 4 dimensional spacetime metric is decomposed into a spatial metric on a 3 dimensional spatial hypersurface, $\gamma_{ij}$, and an extrinsic curvature $K_{ij}$, which are both evolved along a chosen local time coordinate $t$. Since one is free to choose what is space and what is time, the gauge choice must also be specified. 
The line element of the decomposition is
\begin{equation}
ds^2=-\alpha^2\,dt^2+\gamma_{ij}(dx^i + \beta^i\,dt)(dx^j + \beta^j\,dt)\,,
\end{equation}
where $\alpha$ and $\beta^i$ are the lapse and shift, the gauge parameters.  These parameters are specified on the initial hypersurface (see below) and then allowed to evolve using gauge-driver equations, in accordance with the puncture gauge \cite{Campanelli:2005dd}\cite{Baker:2005vv}, for which the evolution equations are
\begin{align} \label{eqn:MovingPuncture}
&\partial_t \alpha = - \mu \alpha K + \beta^i \partial_i \alpha ~ , \\
&\partial_t \beta^i = B^i ~ , \\
&\partial_t B^i = \frac{3}{4} \partial_t \Gamma^i - \eta B^i ~ ,
\end{align}
where the constants $\eta$, of order $1/M_{ADM}$, and $\mu$, of order 1, may be varied by the user to improve stability. The effect of the moving puncture gauge is to avoid resolving the central singularity of any black hole that may form. It was shown that in this gauge the central gridpoints asymptote to a fixed radius within the event horizon, the so-called ``trumpet'' solution described in \cite{Hannam:2008sg}. Thus explicit numerical excision of the central singularity is not required. While constraint violation may occur at the central point due to taking gradients across the puncture, these remain within the horizon and do not propagate into the outside spacetime.

\subsection{Evolution equations}

In $\textsc{GRChombo}$ the induced metric is decomposed as 
\begin{equation}
\gamma_{ij}=\frac{1}{\chi}\,\tilde\gamma_{ij} \quad \det\tilde\gamma_{ij}=1 \quad \chi = \left(\det\gamma_{ij}\right)^{-\frac{1}{3}}  ~ .
\end{equation}
The extrinsic curvature is decomposed into its trace, $K=\gamma^{ij}\,K_{ij}$, and its traceless part $\tilde\gamma^{ij}\,\tilde A_{ij}=0$ as
\begin{equation}
K_{ij}=\frac{1}{\chi}\left(\tilde A_{ij} + \frac{1}{3}\,K\,\tilde\gamma_{ij}\right) ~ .
\end{equation}
The conformal connections $\tilde\Gamma^i=\tilde\gamma^{jk}\,\tilde\Gamma^i_{~jk}$ where $\tilde\Gamma^i_{~jk}$ are the Christoffel symbols associated with the conformal metric $\tilde\gamma_{ij}$.

\noindent The evolution equations for BSSN are then
\begin{align}
&\partial_t\chi=\frac{2}{3}\,\alpha\,\chi\, K - \frac{2}{3}\,\chi \,\partial_k \beta^k + \beta^k\,\partial_k \chi ~ , \label{eqn:dtchi2} \\
&\partial_t\tilde\gamma_{ij} =-2\,\alpha\, \tilde A_{ij}+\tilde \gamma_{ik}\,\partial_j\beta^k+\tilde \gamma_{jk}\,\partial_i\beta^k \nonumber \\
&\hspace{1.3cm} -\frac{2}{3}\,\tilde \gamma_{ij}\,\partial_k\beta^k +\beta^k\,\partial_k \tilde \gamma_{ij} ~ , \label{eqn:dttgamma2} \\
&\partial_t K = -\gamma^{ij}D_i D_j \alpha + \alpha\left(\tilde{A}_{ij} \tilde{A}^{ij} + \frac{1}{3} K^2 \right) \nonumber \\
&\hspace{1.3cm} + \beta^i\partial_iK + 4\pi\,\alpha(\rho + S) \label{eqn:dtK2} ~ , \\
&\partial_t\tilde A_{ij} = \left[-D_iD_j \alpha + \chi \alpha\left( R_{ij} - 8\pi\,\alpha \,S_{ij}\right)\right]^\textrm{TF} \nonumber \\
&\hspace{1.3cm} + \alpha (K \tilde A_{ij} - 2 \tilde A_{il}\,\tilde A^l{}_j)  \nonumber \\
&\hspace{1.3cm} + \tilde A_{ik}\, \partial_j\beta^k + \tilde A_{jk}\,\partial_i\beta^k \nonumber \\
&\hspace{1.3cm} -\frac{2}{3}\,\tilde A_{ij}\,\partial_k\beta^k+\beta^k\,\partial_k \tilde A_{ij}\,   \label{eqn:dtAij2} ~, \\ 
&\partial_t \tilde \Gamma^i=2\,\alpha\left(\tilde\Gamma^i_{jk}\,\tilde A^{jk}-\frac{2}{3}\,\tilde\gamma^{ij}\partial_j K - \frac{3}{2}\,\tilde A^{ij}\frac{\partial_j \chi}{\chi}\right) \nonumber \\
&\hspace{1.3cm} -2\,\tilde A^{ij}\,\partial_j \alpha +\beta^k\partial_k \tilde\Gamma^{i} \nonumber \\
&\hspace{1.3cm} +\tilde\gamma^{jk}\partial_j\partial_k \beta^i +\frac{1}{3}\,\tilde\gamma^{ij}\partial_j \partial_k\beta^k \nonumber \\
&\hspace{1.3cm} + \frac{2}{3}\,\tilde\Gamma^i\,\partial_k \beta^k -\tilde\Gamma^k\partial_k \beta^i - 16\pi\,\alpha\,\tilde\gamma^{ij}\,S_j ~ . \label{eqn:dtgamma2}
\end{align} 
The scalar field matter evolution equations are
\begin{align}
&\partial_t \phi = \alpha \Pi_M +\beta^i\partial_i \phi \label{eqn:dtphi2} ~ , \\
&\partial_t \Pi_M=\beta^i\partial_i \Pi_M + \alpha\partial_i\partial^i \phi + \partial_i \phi\partial^i \alpha \\
&\hspace{1.3cm} +\alpha\left(K\Pi_M-\gamma^{ij}\Gamma^k_{ij}\partial_k \phi+\frac{dV}{d\phi}\right) \label{eqn:dtphiM2} ~ ,
\end{align} 
where the second order Klein Gordon equation has been decomposed into two first order equations as is usual. 

The stress energy tensor for a single scalar field is
\begin{equation}
T_{ab} = \nabla_a \phi \nabla_b \phi - \frac{1}{2} g_{ab} (\nabla_c \phi \, \nabla^c \phi + 2V) \ .
\end{equation}
and the various components of the matter stress tensor are calculated from this as
\begin{align}
&\rho = n_a\,n_b\,T^{ab}\,,\quad S_i = -\gamma_{ia}\,n_b\,T^{ab}\,, \nonumber \\
&S_{ij} = \gamma_{ia}\,\gamma_{jb}\,T^{ab}\,,\quad S = \gamma^{ij}\,S_{ij} \, .
\label{eq:Mattereqns}
\end{align}

\noindent The Hamiltonian constraint is
\begin{equation}
\mathcal{H} = R + K^2-K_{ij}K^{ij}-16\pi \rho \, . \label{eqn:Ham}
\end{equation}
\noindent The momentum constraint is
\begin{equation}
\mathcal{M}_i = D^j (K_{ij} - \gamma_{ij} K) - 8\pi S_i \, .  \label{eqn:Mom}
\end{equation}

\subsection{Specific notes on the current simulations}

Some further details regarding the simulations in this paper are noted in the section below.

\subsubsection{Shift, lapse and AMR conditions}

The shift is set initially to zero.  The lapse condition was chosen so that the slicing was approximately logarithmic in the scale factor, that is the lapse was driven dynamically towards $\alpha=\ln(a)$ with $a = 1/\chi^2$ the (local) scale factor. The initial conditions for the lapse were set according to this condition based on the relaxed value of $\chi$. This was not essential to the stability of the simulations, but was found to give the most ``smooth'' and efficient results, with certain other choices which were tested resulting in local collapses of the coordinate observers before the maximum number of $e$-folds was reached. While the choice of constant $\alpha=1$ slicing actually gave stable results, it was rather inefficient as the timesteps did not grow in line with the scale factor. On the other hand approximately conformal time ($\alpha=a$) tended to result in instabilities developing.

For the majority of the simulation, during the inflationary period, a fixed grid was sufficient. Regridding was triggered by high gradients of $\rho_{A_{ij}}$, which was seen in the cases of gravitational collapse of the wave background (and just below where the values almost collapsed and then recovered), and by high gradients in $\phi$, when the value fell towards the reheating minimum $\phi_*$ at the end of inflation.

\subsubsection{Meaurement of $e$-folds}

We used the local value of $\chi$ at the first point of failure of inflation to calculate the number of $e$-folds achieved, and which are displayed in the figures. Failure was defined by the value of $\phi$ reaching the reheating minimum $\phi_*$ at some point in the spacetime, and it was at this point that the local number of $e$-folds was measured. As in our previous work, we see that the rest of the spacetime is subsequently dragged down into the minimum by the gradients formed. An alternative would have been to calculate the average over the spatial slice at the point of failure, although this would be more sensitive to the gauge/slicing conditions, or to wait for the entire spacetime to be dragged into the minimum before measuring, but this was computationally more expensive. The exact measure of $e$-folds, local or average across the grid, should not affect the results obtained, except to change the exact numerical values, which in any case are strongly model dependent.

\begin{figure}
\begin{center}
\includegraphics[width=8cm]{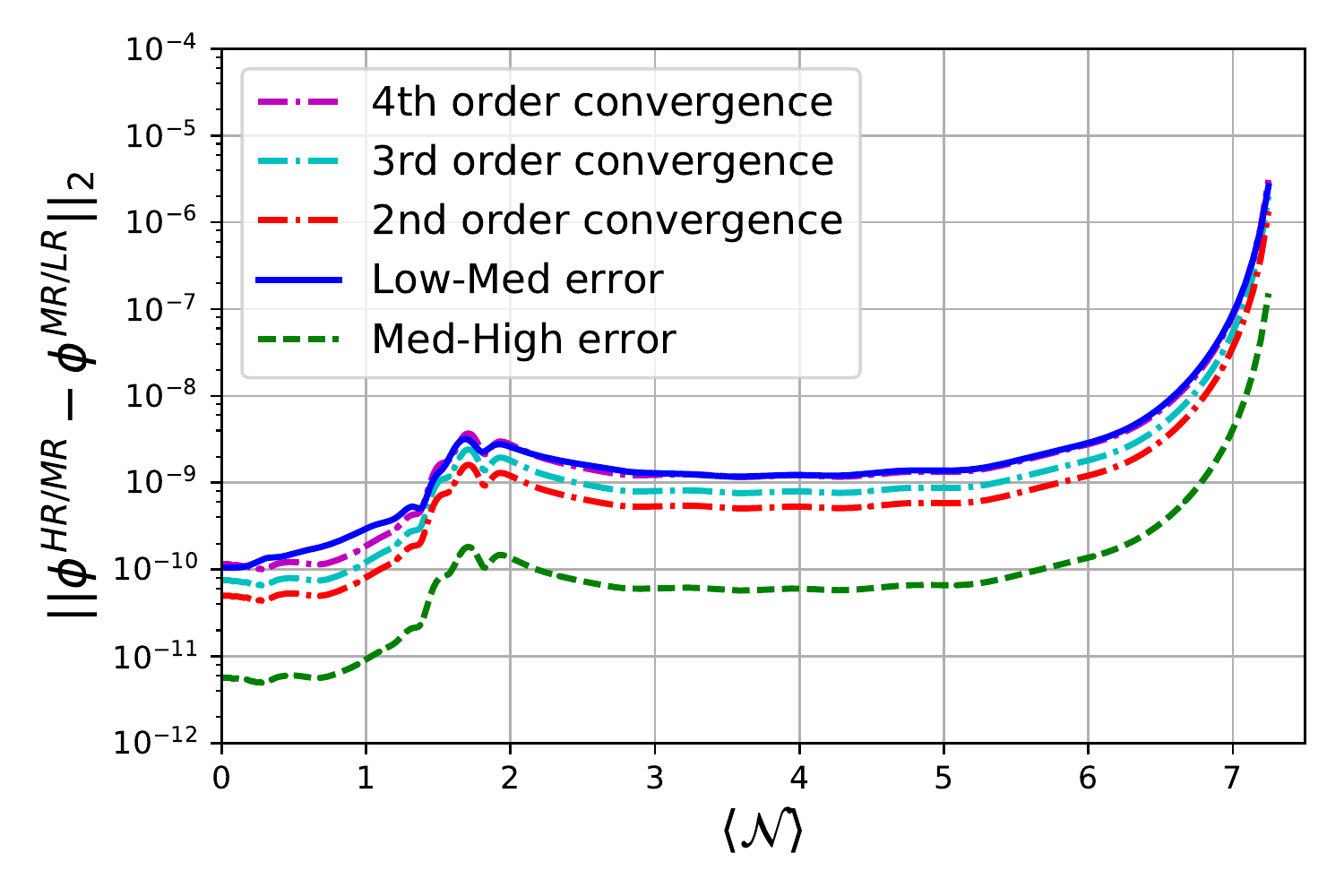}
\includegraphics[width=8cm]{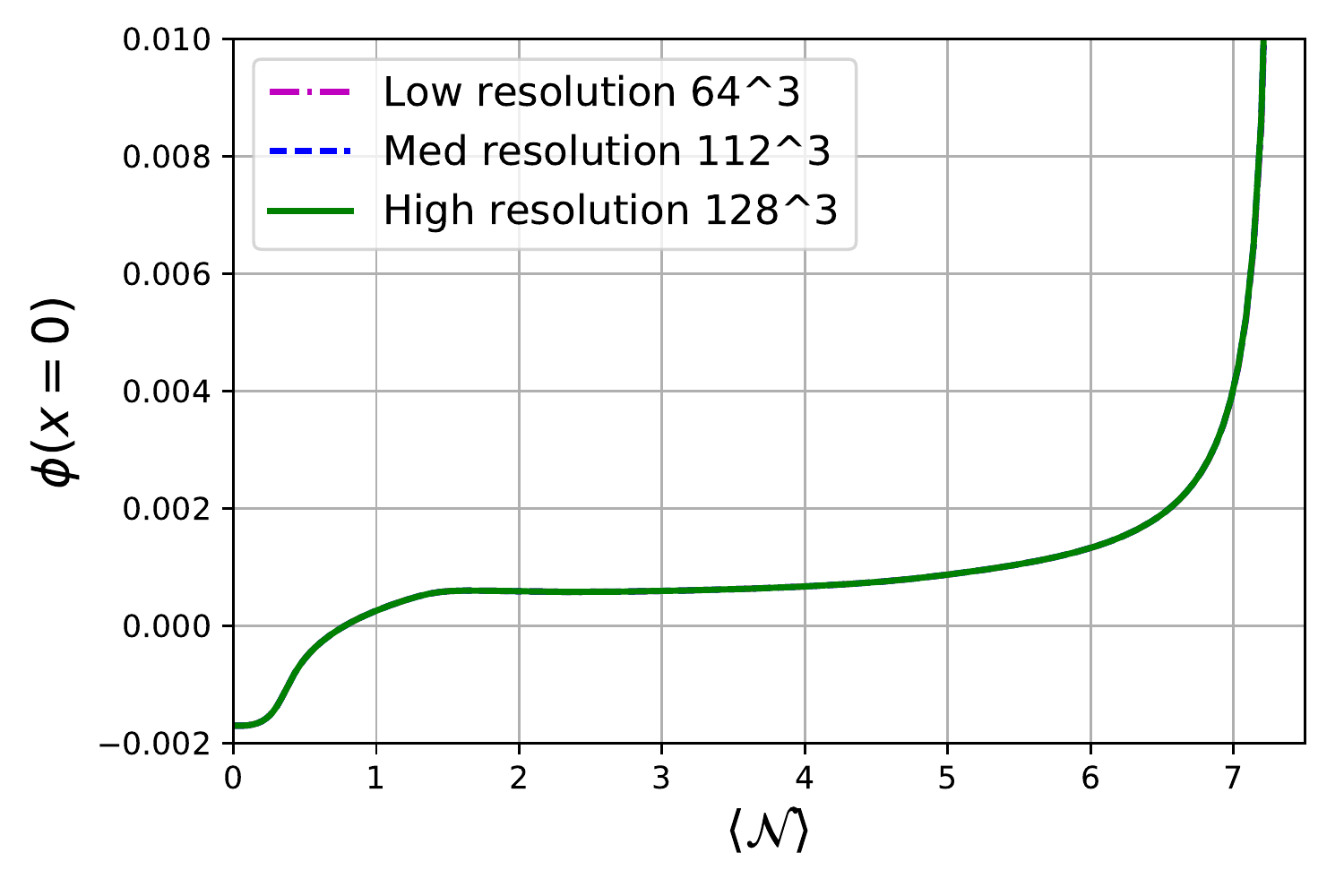}
\caption{A convergence test which compares the difference in the values of $\phi$ for the simulations for which $\Delta \phi=0.0006 \mpl$ and  $\Delta A= 1.5 \times  10^{-11} \mpl$. For these tests the resolution was fixed, and set to $64^3$, $112^3$ and $128^3$ in the low resolution (LR), medium resolution (MR) and high resolution (HR) runs respectively. The differences are taken at 125 different points on the grid and the L2 norm taken of all the values. The HR-MR and MR-LR cases are compared to check the order of convergence. The reduction in the error in the MR-HR case is consistent with 4th order convergence. Only the evolution period is shown, since the values are fixed for the relaxation period.}
\label{fig:Convergence_phi}
\end{center}
\end{figure}

\subsubsection{Convergence and constraint violation} \label{iteration}

The results of an example convergence test for the simulation ($\Delta \phi=0.0006 \mpl$,  $\Delta A= 1.5 \times  10^{-11} \mpl$) is shown in figure \ref{fig:Convergence_phi} for the field value. The tests used a fixed mesh with the resolution set to $64^3$, $112^3$ and $128^3$ respectively in the low resolution, medium resolution and high resolution runs. The plots indicates that level of convergence is approximately 4th order for most of the simulation. In the simulations for which we present our results we used the highest resolution as our coarsest grid, although we permitted regridding with the full AMR capabilities of the code. In practise the resolution did not often go above the coarsest refinement, except during periods where the tensor fluctuations collapsed, or the field fell into the reheating minimum at the end of the simulations. This would normally reduce the order of convergence to 3rd order or lower. While this has the potential to introduce additional sources of error, we did not see any indication of significant problems in comparison to the fixed mesh case.

\begin{figure}
\begin{center}
\includegraphics[width=8cm]{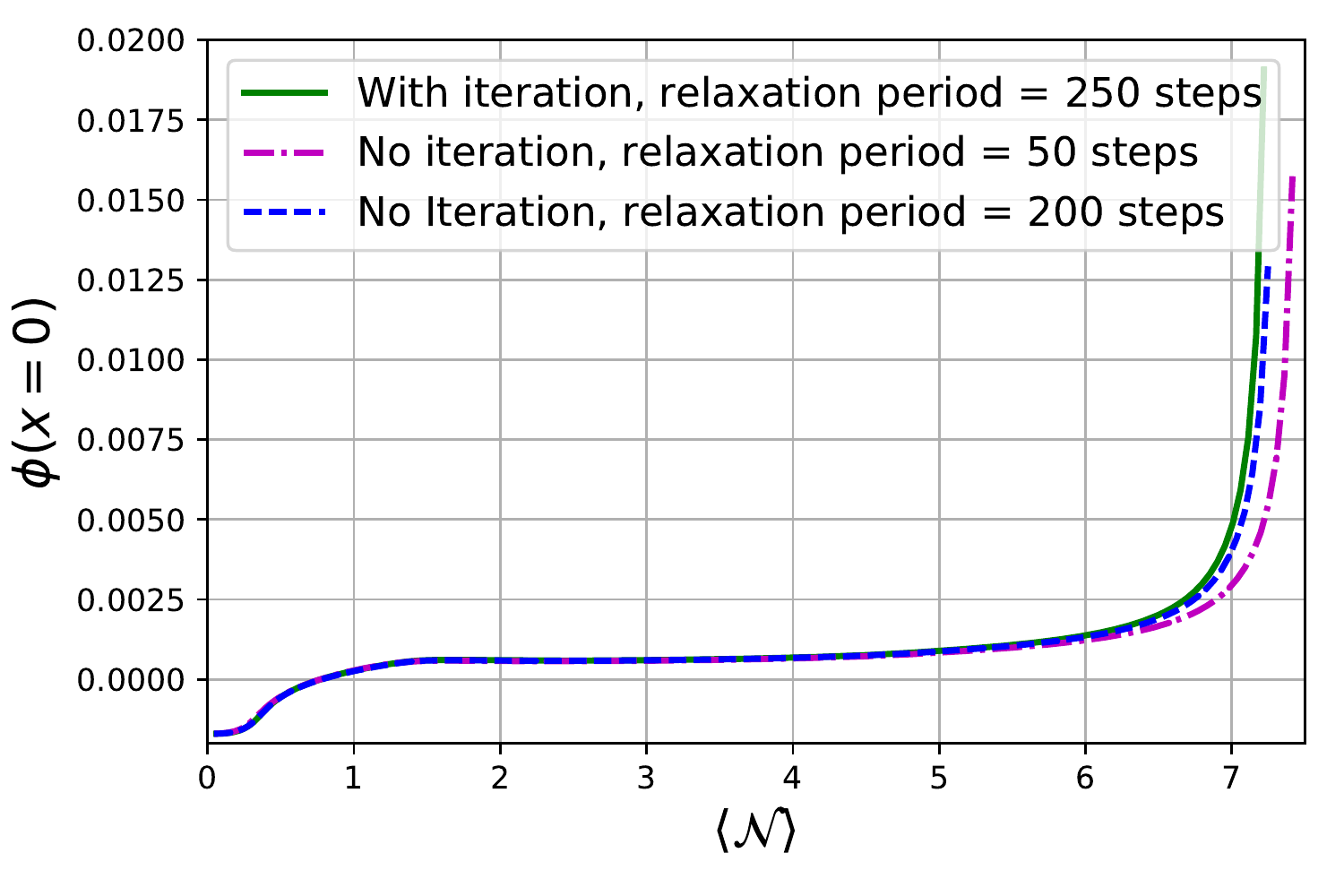}
\includegraphics[width=8cm]{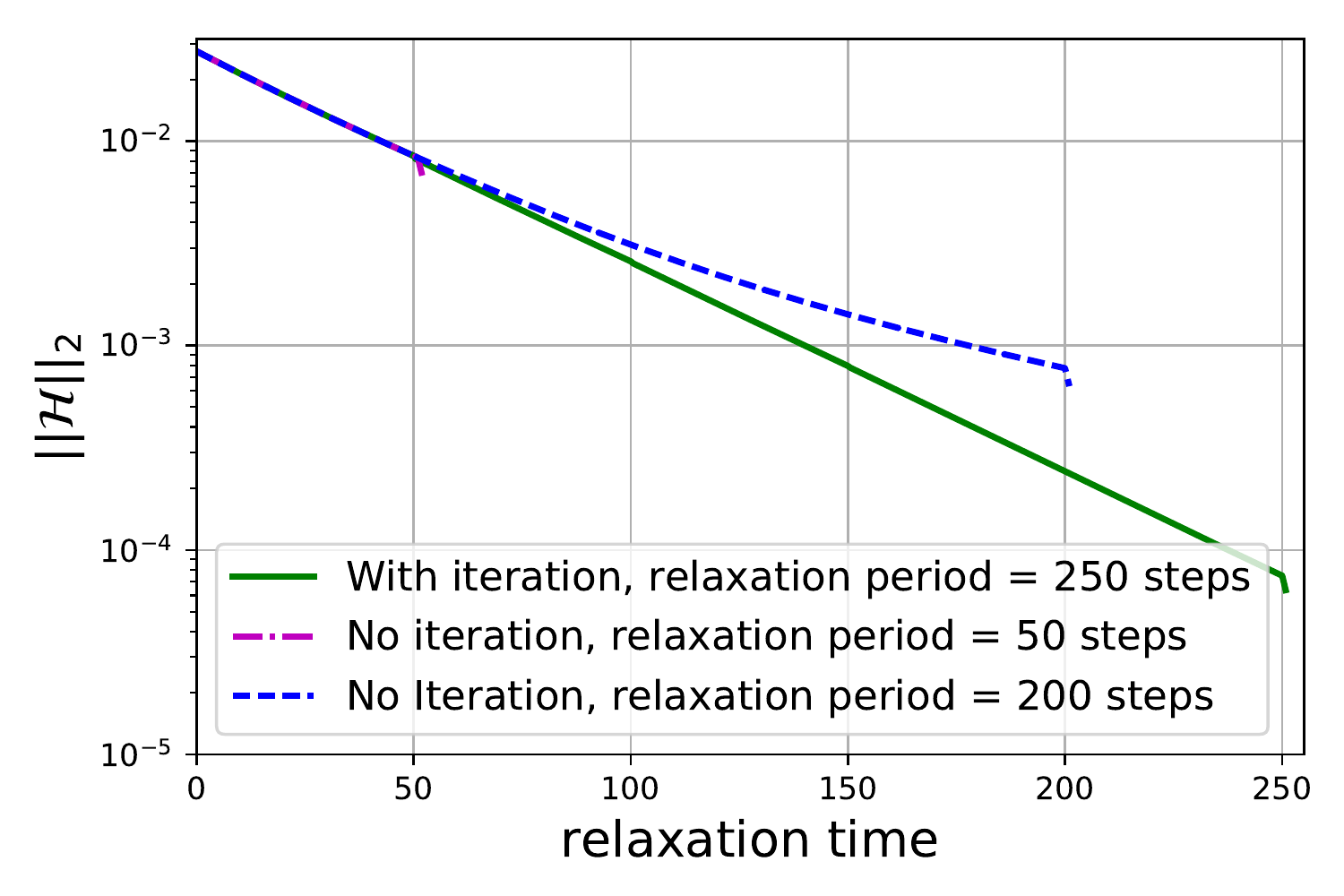}
\includegraphics[width=8cm]{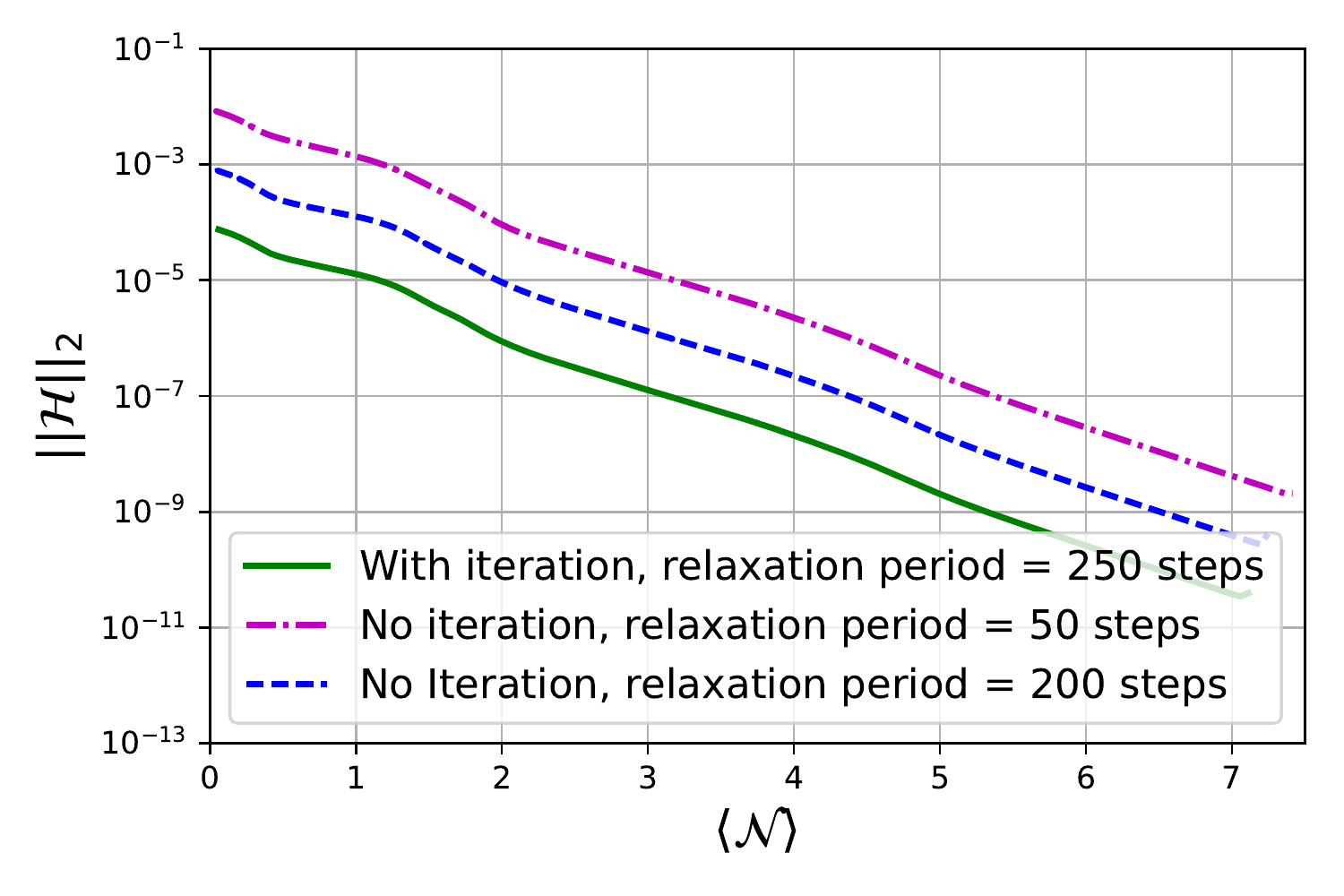}
\caption{The top figure is the evolution of the field $\phi$ as in figure \ref{fig:Convergence_phi}, but shows the effect of a longer or shorter relaxation period, and the iteration or non iteration of the initial value of $K$ to better satisfy the Hamiltonian constraint. The second shows the evolution of the Hamiltonain constraint during the relaxation period and the third the Hamiltonian constraint during the main evolution. The blue dashed line corresponds to the case used in the simulations presented in the main text - a relaxation period of 200 timesteps but no iteration of $K$. The green line is an order of magnitude better, being obtained using an iteration of the value of $K$ and a longer overall relaxation period. For comparison, a shorter period of only 50 timesteps is shown as a pink dot-dashed line, which gives an order of magnitude worse constraint violation. We show that obtaining a smaller constraint violation, either by a longer relaxation period, or by iterating the estimate of $K$, does not significantly change our results, and that the Hamiltonian constraint remains stable and bounded in all cases.}
\label{fig:Convergence_Ham}
\end{center}
\end{figure}

Figure \ref{fig:Convergence_Ham} shows detail relating to the constraint violation as a result of the relaxation procedure. During the relaxation period the Hamiltonian constraint should converge towards zero (by adjusting the spatial profile of $\chi$), but in fact relaxes towards a small non zero value if no iteration of the initial value of $K$ is performed. This error is the dominant contribution to the constraint violation during the evolution, (above numerical truncation errors), but remains stable and bounded throughout. It arises from setting $K^2 = \langle 3/2 \tilde A_{ij} \tilde A^{ij} + 24 \pi \rho \rangle$, assuming that $\chi=1$ everywhere in the calculation of $\tilde A_{ij}$, whereas we then solve for a spatially varying $\chi$. As noted in footnote 4, in order to remove this error we would need to re-solve for the correct value of $K$ with the new profile for $\chi$, and then iterate the procedure. We show in the figure that the iteration makes no significant difference to the outcome of the results compared to the case used in the simulations, where a relaxation period of 200 timesteps was used but no iteration of the value of $K$. Compared to the iterated case, the difference in both $K$ and the measured number of e-folds to the end of inflation were less than 1\%, which is well below the accuracy to which they are presented in our results. However, it is clear that an iterated relaxation does improve the accuracy of the results, and thus would be required for more accurate work in, for example, cosmological parameter estimation.

\bibliography{inhomoinf2.bib}

\end{document}